\documentclass[times,12pt]{article}

\usepackage{times}
\usepackage{natbib}
\usepackage{graphicx}
\usepackage{amsmath}
\usepackage{amsthm}
\usepackage{amssymb}
\usepackage{setspace}
\usepackage[margin=1in]{geometry}
\RequirePackage[colorlinks,citecolor=blue,urlcolor=blue]{hyperref}
\usepackage{breakurl}
\usepackage{bbm}
\usepackage{multirow}

\DeclareMathOperator{\logit}{logit}
\DeclareMathOperator{\Var}{Var}

\newcommand{\bfbeta}{{\boldsymbol \beta}}
\newcommand{\bftheta}{{\boldsymbol \theta}}

\begin{document}

\title{Quantifying the effect of interannual ocean variability on the attribution of extreme climate events to human influence}

\author{ Mark D. Risser$^{\> a \>*}$, D\'aith\'i A. Stone$^{\>c}$, Christopher J. Paciorek$^{\> b}$,\\ Michael F. Wehner$^{\>c}$, Oliver Ang\'elil$^{\>d}$}
\date{}
\maketitle  

{\scriptsize
\hskip4ex $^a$ Earth and Environmental Sciences Division, Lawrence Berkeley National Laboratory, Berkeley, CA, USA

\hskip4ex $^b$ Department of Statistics, University of California, Berkeley, Berkeley, CA, USA

\hskip4ex $^c$ Computational Research Division, Lawrence Berkeley National Laboratory, Berkeley, CA, USA

\hskip4ex $^d$ Climate Change Research Centre, University of New South Wales, Sydney, Australia

\hskip4ex $^*$ Correspondence email: \tt{mdrisser@lbl.gov}
}

\vskip3ex

\begin{abstract}
In recent years, the climate change research community has become highly interested in describing the anthropogenic influence on extreme weather events, commonly termed ``event attribution.'' Limitations in the observational record and in computational resources motivate the use of uncoupled, atmosphere/land-only climate models with prescribed ocean conditions run over a short period, leading up to and including an event of interest. In this approach, large ensembles of high-resolution simulations can be generated under factual observed conditions and counterfactual conditions that might have been observed in the absence of human interference; these can be used to estimate the change in probability of the given event due to anthropogenic influence.  However, using a prescribed ocean state ignores the possibility that estimates of attributable risk might be a function of the ocean state. Thus, the uncertainty in attributable risk is likely underestimated, implying an over-confidence in anthropogenic influence.

In this work, we estimate the year-to-year variability in calculations of the anthropogenic contribution to extreme weather based on large ensembles of atmospheric model simulations. Our results both quantify the magnitude of year-to-year variability and categorize the degree to which conclusions of attributable risk are qualitatively affected. The methodology is illustrated by exploring extreme temperature and precipitation events for the northwest coast of South America and northern-central Siberia; we also provides results for regions around the globe. While it remains preferable to perform a full multi-year analysis, the results presented here can serve as an indication of where and when attribution researchers should be concerned about the use of atmosphere-only simulations.
\end{abstract}

\doublespacing


\section{Introduction} \label{introduction}

In recent years, public interest has motivated the climate change research community to become highly interested in describing the influence of anthropogenic emissions on extreme weather events, commonly termed ``event attribution'' \citep{NAS_2016}.  Limitations in the observational record motivate the use of climate models to estimate anthropogenic influence and risk: first, extremes are (by definition) rare, meaning their occurrence in the observational record is sparse; second, no observations at all are available for a counterfactual world without human influence.  The use of observations from a century ago has issues both in terms of representing anthropogenic influence and of data quality. Instead, climate models can be used to simulate the weather conditions in both a factual world driven by observed boundary conditions (e.g., greenhouse gas concentrations, solar luminosity) and a counterfactual world driven by what those boundary conditions might have been in the absence of historical human influence on the climate. However, extreme weather is often localized, which requires these models to be run at a resolution fine enough to resolve the physics in the model that generate extreme weather. And, for a probabilistic approach examining rare events, large ensembles of simulations are needed for both scenarios.

This leads to a computational problem: fully-coupled, high resolution models that collectively model land, ocean, and atmospheric processes come with an extremely high computational cost. Fully-coupled models have long memory, meaning that it can take decades to centuries of model time to move away from biases introduced by the initial conditions. 
An additional issue is that significant biases can arise in full models of the climate system, wherein biases in ocean conditions can induce biases in the atmospheric state, which in turn may reinforce the ocean biases.

To navigate these difficulties when considering atmospheric extremes, an alternative approach involves using uncoupled atmosphere/land-only models with prescribed ocean conditions.  The boundary conditions for both the factual and counterfactual worlds now include the observed ocean state, with the counterfactual values adjusted according to a spatially and seasonally evolving estimate of the warming due to anthropogenic emissions.  These models have less spin-up time, and large ensembles can be obtained much more quickly. Additionally, fixing the ocean conditions can provide some benefit in that it removes dependence on the quality of the ocean model's simulation of low frequency processes like El Ni\~no and on biases in the ocean model's ocean surface state.

Of course, this approach involves some important assumptions relating to the ocean surface state and the way the atmosphere interacts with the ocean.  
While researchers may be interested in the conditional case (e.g. during El~Ni\~no events),
when the results are interpreted as a general overall assessment of the role of anthropogenic emissions, these conditions become important assumptions. All attribution analyses involve some elements of uncertainty (\citealp{Jeon2016,HansenG_AuffhammerM_SolowAR_2014,PallP_AinaT_etalii_2011}), but particular types of uncertainty are specific to the atmosphere/land-only modeling approach. 
%
First and foremost, the strategy of fixing the sea surface temperatures (SSTs) does not account for the internal variability of the ocean system.  This involves three important assumptions.
\begin{enumerate}
  \item \textbf{Invariance to ocean state:}  The anthropogenic influence on extreme weather is assumed to be independent of the ocean state, so for instance anthropogenic influence is considered identical during El~Ni\~no events as during La~Ni\~na events, ignoring the possibility that the atmospheric response to these two types of events may have different sensitivities to anthropogenic forcing.
  \item \textbf{No change in ocean variability: } The nature of ocean variability, for instance the frequency and spatial structure of El~Ni\~no events, is assumed to be unaffected by anthropogenic emissions, beyond any spatial or seasonal gradients imposed by the attributable ocean warming estimate.
  \item \textbf{Unimportance of atmosphere-ocean interaction: } Short time-scale atmosphere-ocean interactions, for instance that might occur under tropical cyclones, are assumed to be unimportant.
\end{enumerate}
This paper specifically examines the accuracy of the first assumption.

The impact of ocean variability on atmosphere-model-based event attribution assessments has yet to be quantified.  Identifying the effect of oceanic variability boils down to investigating how the experimental design of a single-year, atmosphere-only attribution framework impacts attribution statements. For example, attribution statements may be highly sensitive to the specific temporal period used for the study, over periods short enough that a long-term trend is unimportant (\citealp{OttoFEL_BoydE_etalii_2015}; \citealp{WolskiP_StoneD_etalii_2014}). These sensitivities might vary for different events and different regions in the world, and thus it would be helpful to identify events and regions of the world for which atmospheric model designs are adequate for reasonably accurate event attribution assessment.

In this work, we develop and implement a hierarchical statistical model that allows us to address the uncertainty introduced in attribution studies from the use of uncoupled, atmosphere/land-only model simulations (in addition to sampling uncertainty from the limited number of simulations). Specifically, this framework allows us to quantify the effect of ocean variability on statements of risk, while incorporating and adjusting for long-term trends. Furthermore, the model is fit using a Bayesian framework, which allows for the development of a diagnostic tool for identifying event types and regions of the globe for which single-year, atmosphere-only climate simulations may be sufficient for assessing the anthropogenic contribution to risk. Finally, we present a framework within which future single-year, atmosphere-only event attribution studies can assess the robustness of their results to the effect of ocean variability. 

The paper is organized as follows. In Section \ref{app_fmwk}, we introduce our event attribution framework and describe the specific climate model simulations used in this analysis. In Section \ref{methods}, we outline the hierarchical statistical model, including a discussion of the implied model for the risk ratio. Section \ref{diag1} introduces several diagnostic tools for applying estimates of the effect of ocean variability in a general event attribution study. In Section \ref{results}, we apply the statistical model to two case studies, and in Section \ref{global} we describe and apply the diagnostic tool generally to several dozen regions around the world. Section \ref{discussion} concludes the paper.

\section{Attribution framework} \label{app_fmwk}

\subsection{Climate model simulations}

We examine the frequency of extreme monthly values of temperature and precipitation in two ensembles of simulations of the CAM5.1 global atmosphere/land climate model.  The model is run in its conventional $\sim 1^{\circ}$ longitude/latitude configuration \citep{NealeRB_ChenC-C_etalii_2012}.  Simulations have been run under the experiment protocols of the C20C+ Detection and Attribution Project \citep{StoneDA_PallP_2016}, following two historical scenarios \citep{AngelilO_StoneD_etalii_2016}.  The first set of simulations (ALL) is driven by observed boundary conditions of atmospheric chemistry (greenhouse gases, tropospheric and stratospheric aerosols, ozone), solar luminosity, land use/cover, and the ocean surface (temperature and ice coverage).  The second set of simulations (NAT) is driven by what observed boundary conditions might have been in the absence of historical anthropogenic emissions:  the anthropogenic component of atmospheric chemistry is set to year-1855 values, ocean temperatures are cooled by a seasonally- and spatially-varying estimate of the warming attributable to anthropogenic emissions 
(that is the same across all simulations),
 and sea ice concentrations are adjusted for consistency with the ocean temperatures \citep{StoneDA_PallP_2016}.  Simulations within a scenario differ only in the starting conditions.  Here we examine data during the 1982--2013 period; while simulations are available for some earlier years, we want to ensure that incorporation of satellite data into the sea surface temperature product used by the simulations in 1981 does not influence our conclusions \citep{HurrellJW_HackJJ_etalii_2008}. 
Furthermore, this provides a period in which risk ratios should be of the same order of magnitude across years (even if exhibiting a substantial long-term trend), which would not be the case as one goes further back in time. In this analysis, we make this assumption for second order terms but not for first order terms.
For each scenario we have 50 simulations during 1982-1996, with an additional 50 (hence 100 total) during 1997-2010, and an additional 300 (400 total) during 2011-2013 (Table~\ref{ensSize}).  The data and further details on the simulations are available at {\url{http://portal.nersc.gov/c20c}}.

\begin{table}[!t]
\caption{Ensemble sizes for the CAM5.1 simulations from 1982-2013.}
\begin{center}
\begin{tabular}{|c|c|c}
\hline
\textbf{Year range ($t$)} & \textbf{Ensemble size ($n_t$)} \\ \hline \hline
1982--1996 & 50 \\ \hline
1997--2010 & 100 \\ \hline
2011--2013 & 400 \\
\hline
\end{tabular}
\end{center}
\label{ensSize}
\end{table}%

\subsection{Using the risk ratio for attributional statements}

In order to make attribution statements for the impact of anthropogenic influences on the occurrence of extreme events, we follow a probabilistic extreme event attribution approach (\citealp{AllenM_2003,StoneDA_AllenMR_2005a,HansenG_AuffhammerM_SolowAR_2014}), by considering occurrence probabilities $p_A$ and $p_N$. These quantities represent the probability that a predefined event will occur over a predefined spatial and temporal domain, with $p_A$ representing the observed world or ``world as it is'' (henceforth the so-called ALL forcings scenario), including both natural and anthropogenic influences on the climate, and $p_N$ representing the counterfactual scenario of the ``world as it may have been'' (henceforth the so-called NAT forcings scenario) without anthropogenic effects. We consider the ratio of these probabilities, or the risk ratio
\[
RR = \frac{p_A}{p_N},
\]
which mirrors the measurement of risk used in epidemiology and environmental law \citep{StoneDA_AllenMR_2005a} and has a number of desirable properties \citep{NAS_2016}. A risk ratio of larger than one indicates that anthropogenic influences have increased the likelihood of an event's occurrence, while a risk ratio of less than one indicates a decreased likelihood of occurrence.

In this work, we are interested in estimating the risk ratio over time, at a total of $T$ time increments. In what follows, the time increments represent years, but this framework could similarly be used for months, seasons, or decades, for example. Therefore, we must consider 
\[
RR_t = \frac{p_{At}}{p_{Nt}}  \hskip5ex t = 1, \dots, T,
\]
where $p_{At}$ and $p_{Nt}$ are the occurrence probabilities for a predefined time increment $t$. 

\subsection{Modeling probabilities and event definition}

In general, while there are a variety of approaches for estimating $p_A$ and $p_N$, two popular methods are the nonparametric (or binomial) approach and extreme value analysis. The binomial approach is desirable in that it makes no assumptions about the underlying distribution of the variable of interest (e.g., precipitation or temperature), however it becomes essentially useless when the event of interest is extremely rare or unobserved. Extreme value analysis, on the other hand, requires additional assumptions on the behavior of the variable of interest but is able to estimate the probability of extremely rare events, even those that do not occur frequently or at all in an ensemble or the observational record.

In what follows, we considered events of a pre-specified magnitude, and simply used an empirical quantile to define a threshold for determining whether or not a particular event is ``extreme.'' Furthermore, for simplicity, we chose a quantile that is extreme but not too extreme: namely, the one in ten year event. In this case, this nonparametric binomial approach is sufficient for modeling probabilities.

\section{Hierarchical statistical model for ALL and NAT probabilities} \label{methods}

Unlike other approaches that have explored estimation and uncertainty quantification for individual time points or years \citep{PallP_AinaT_etalii_2011}, recall that our goal is to make systematic statements about the behavior of the risk ratio for multiple time points. In doing so, we can make statements about the effect of various explanatory variables on event probabilities and the risk ratio in addition to quantifying interannual variability. A hierarchical statistical modeling framework allows us to address each of these questions by borrowing information over time.

Using a nonparametric approach for modeling the probabilities, the first level of the hierarchical model relates  the climate model simulations to the occurrence probabilities. While we are interested in modeling the risk ratio across different years, in order to account for seasonality we define monthly random variables 
\begin{equation} \label{zDefs}
\begin{array}{rcl} 
Z_{Atj} & = & \text{ number of ALL simulations in month $j$ of year $t$ in which the event occurred} \\ 
Z_{Ntj} & = & \text{ number of NAT simulations in month $j$ of year $t$ in which the event occurred;} \\ 
\end{array}
\end{equation}
each of these random variables is modeled in a statistical framework as arising from a binomial experiment with $n_t$ total trials and corresponding success probabilities $p_{ktj}$, for $k \in \{A,N\}$ and $t=1, \dots, T$. Furthermore, conditional on the probabilities ${\bf p} = \{ p_{ktj}: k \in \{A,N\}; t=1, \dots, T; j = 1, \dots, 12 \}$, the random variables ${\bf Z} = \{ Z_{ktj}:  k \in \{A,N\}; t=1, \dots, T; j = 1, \dots, 12 \}$ are independent. In other words, the joint probability mass function (pmf) of ${\bf Z}$ conditional on ${\bf p}$ is
\begin{equation} \label{likelihood}
p({\bf Z} | {\bf p}) = \prod_{k \in \{A, N\}} \prod_{t=1}^T \prod_{j=1}^{12} p(Z_{ktj} ; p_{ktj}),
\end{equation}
where the marginal pmfs $p(Z_{ktj}; p_{ktj})$ are binomial.

The second level of the hierarchy consists of a model that allows the unknown parameters to vary based on covariate information, while also tying together parameters that correspond to the same time period (i.e., $p_{Atj}$ and $p_{Ntj}$ for common $t$ and $j$). For this level, we used
\begin{equation} \label{full_model}
\logit p_{ktj} \equiv \log \left( \frac{p_{ktj}}{1-p_{ktj}} \right) =  {\bf x}_{kt}^\top\bfbeta_k + \alpha_t + \delta_t \mathbbm{1}_{\{k=A\}} + \gamma_j ,
\end{equation}
for $k \in \{A,N\}$, $t=1, \dots, T$, and $j = 1, \dots, 12$. Here, ${\bf x}_{kt} = (1, x_{kt1}, \dots, x_{ktp})$ is a vector of covariates for scenario $k \in \{A,N\}$, $\bfbeta_k = (\beta_{k0}, \dots, \beta_{kp})$ is a scenario-specific vector of unknown regression coefficients, and $\mathbbm{1}_{\{\cdot\}}$ is an indicator function. The strategy in (\ref{full_model}) is a mixed effects logistic regression, a special case of the more general statistical approach of generalized linear mixed models (see, e.g., \citealp{glmm}). Intuitively, including the scenario-specific covariates ${\bf x}_{kt}$ accounts for any long-term trends present in $p_N$ and $p_A$.

The final level of the hierarchy ties together the year-specific effects $\{ \alpha_t, \delta_t : t= 1, \dots, T\}$ and month-specific effects $\{ \gamma_j : j= 1, \dots, 12\}$ in order to allow for borrowing of information across years and months (called ``partial pooling''). This level of the model specifies
\begin{equation} \label{level3}
\alpha_t \stackrel{\text{iid}}{\sim} N(0, \tau^2), \hskip3ex  \delta_{t} \stackrel{\text{iid}}{\sim} N(0, \sigma^2), \hskip2ex \text{and} \hskip2ex  \gamma_{j} \stackrel{\text{iid}}{\sim} N(0, \omega^2)
\end{equation}
where ``iid'' stands for ``independent and identically distributed'' and $N(a,b)$ denotes a univariate Gaussian (normal) distribution with mean $a$ and variance $b$. In terms of the probabilities, the $\alpha_t$ terms represent yearly deviations in event probabilities (above and beyond deviations due to the long-term trend) that are common to both ALL and NAT scenarios (e.g., El Ni\~no or La Ni\~na events), while the $\delta_t$ terms represent deviations above and beyond the $\alpha_t$ that are specific to the ALL scenario (for example, the magnitude or probability of an event for ALL in an El Ni\~no or La Ni\~na year may be different than for NAT). These $\delta_t$ terms have a specific importance for how the risk ratio is modeled, and will be explored further in Section \ref{rrm}. Finally, the $\gamma_j$ terms represent within-year deviations from the average yearly probabilities, which account for seasonality: for example, in the extratropics of the northern hemisphere, the probability of a hot extreme is essentially zero from October to March, while a hot extreme might be relatively common in July. Therefore, $\gamma_{12}$ (December) would likely be a large negative number (corresponding to smaller probabilities on the $\logit$ scale) while $\gamma_{7}$ (July) would likely be a large positive number (corresponding to larger probabilities on the $\logit$ scale). 

Several notes should be made regarding the statistical modeling of the monthly effects. First, because the population for the $\gamma_j$ terms represents only 12 months (and because this sample of 12 represents all possible months), we impose the restriction that $\sum_{j=1}^{12} \gamma_j = 0$ to ensure that the sample average is equal to the population average; also, this constraint ensures we can interpret the $\delta_t$ terms directly with respect to the yearly risk ratio (see Section \ref{rrm}). Second, note that the $\gamma_j$ terms depend on neither $k$ (the scenario) nor $t$ (the year); the implication is that we impose the same seasonality on the simulations in both ALL and NAT and a constant seasonality over time. While this may be a restrictive assumption, we make this choice for several reasons. First, it simplifies the parameter interpretations and allows us to use a single parameter to capture the variability in the risk ratio over time (again see Section \ref{rrm}). Also, where the annual cycle in probability has a strong seasonal peak, it is only really the values for that season that matter. Finally, note that the exchangeable model for the $\gamma_j$ terms in (\ref{level3}) does not require that the seasonality be modeled smoothly in time. We acknowledge that our approach is somewhat simplistic, but in doing so we avoid having to customize the model to account for the differing seasonalities of each region and each event type. Instead, the simplified approach of (\ref{level3}) is highly flexible and can automatically adjust to the type of seasonality present for a particular event type and region (accounting for both the length of seasonality and the magnitude of differences across seasons).

The partial pooling of (\ref{level3}) is in contrast to \textit{complete pooling}, which would fix each $\alpha_t \equiv \alpha$, $\delta_t \equiv \delta$, and $\gamma_j \equiv \gamma$, and \textit{no pooling}, which would allow each $\alpha_t$, $\delta_t$, and $\gamma_j$ to be estimated independently of all others. Partial pooling, a popular strategy in random effects modeling (\citealp{GelmanBDA}) and meta analysis (\citealp{Dersimonian1986}), offers a compromise between these two by allowing each year (or month) to have its own effect (i.e., unique $\alpha_t$, $\delta_t$, and $\gamma_j$) while borrowing strength across years (or months) by requiring these effects to come from a common distribution. 
The partial pooling done here is a standard form of statistical
shrinkage that seeks to increase the signal to noise ratio by using
the data from all years to help inform the time-point specific
effects, with the degree to which information is shared across time
points controlled by the data itself.

Up until this point, the model has been presented in general form; in this application, the model was fit within a Bayesian paradigm. In a Bayesian analysis, the unknown parameters are assumed to be random variables, and hence must be assigned a prior distribution. This distribution summarizes all knowledge about the unknown parameters \textit{a priori}, or prior to observing data. For notational simplicity, set $\bftheta = (\boldsymbol{\alpha}, \boldsymbol{\delta}, \boldsymbol{\gamma}, \bfbeta_A, \bfbeta_N, \tau^2, \sigma^2, \omega^2)$; denote the prior as $p(\bftheta)$, and rewrite the likelihood (\ref{likelihood}) in terms of these parameters, i.e, $p({\bf Z} | \bftheta)$. Bayes' Theorem supplies the required machinery to update the prior distribution with observed data (here, ${\bf Z}$) to arrive at the posterior distribution:
\begin{equation} \label{posterior}
p(\bftheta | {\bf Z}) = \frac{p({\bf Z} | \bftheta) p(\bftheta)}{ \int_\bftheta p({\bf Z} | \bftheta) p(\bftheta) d\bftheta},
\end{equation}
i.e., the updated knowledge about $\bftheta$ after observing ${\bf Z}$. While the posterior (\ref{posterior}) is not available in closed form (due to the intractable integral in the denominator of (\ref{posterior}); this is often the case), Markov chain Monte Carlo (MCMC) methods can be used to obtain joint samples from the posterior distribution of $\bftheta$, upon which all subsequent inference is based.

While one advantage of using Bayesian methods (in general) is that any known prior information relating to the parameters can be incorporated by way of the prior distribution $p(\bftheta)$, we instead chose to use a ``noninformative'' prior distribution, which specifies essentially no information about the parameters of the model (in order to avoid any prior biases).

Additional documentation, including more information on the prior specification and computational details for fitting the hierarchical model, is available in the supplemental materials section online. Furthermore, software for fitting the hierarchical model using the {\tt R} programming language and the data sets used in Section \ref{results} are available for download at \url{http://bitbucket.org/markdrisser/timerr_package}.

\subsection{Risk ratio modeling} \label{rrm}

Now, consider what the three-level model specified by (\ref{likelihood}), (\ref{full_model}), and (\ref{level3}) imply about a model for the risk ratio. While the occurrence probabilities have been defined monthly, we are actually interested in averaging over the seasonality to arrive at yearly probabilities, defined as $p_{kt} = \frac{1}{12} \sum_{j=1}^{12} p_{ktj}$. The reason for doing so is that we are less interested in describing how the risk ratio changes monthly (e.g., how the risk ratio in July, 1995 is different from that of August, 1995) and more interested in how the risk ratio changes over longer time scales (years). The fact that the occurrence probabilities are defined for each month (i.e., $p_{ktj}$) is motivated by the need to account for seasonality.

\sloppypar{
Therefore, the risk ratio for an individual year, averaged over all months (from (\ref{full_model})), is  
\begin{equation} \label{riskratio_calc} 
RR_t = \frac{p_{At}}{p_{Nt}} = \frac{ \frac{1}{12} \sum_{j=1}^{12} \logit^{-1}({\bf x}_{At}^\top\bfbeta_A + \alpha_t + \delta_t + \gamma_j)}{ \frac{1}{12} \sum_{j=1}^{12} \logit^{-1}({\bf x}_{Nt}^\top\bfbeta_N + \alpha_t + \gamma_j)}.
\end{equation}
To understand the implications of (\ref{riskratio_calc}) more clearly, note that in dealing with extreme events, the probabilities are often small (i.e., near zero), in which case $\logit(p) = \log(p) - \log(1-p) \approx \log(p)$ and $\logit^{-1}(x) \approx \exp\{x\}$. Therefore, we can approximate the risk ratio (\ref{riskratio_calc}) as
\begin{equation} \label{approxRR}
\begin{array}{rcl}
{RR}_t & \approx & \frac{ \frac{1}{12} \sum_{j=1}^{12}\exp \left\{ \beta_{A0} + \beta_{A1}x_{At1} + \alpha_t + \delta_t + \gamma_j \right\} }{ \frac{1}{12} \sum_{j=1}^{12}\exp \left\{\beta_{N0} + \beta_{N1}x_{Nt1} + \alpha_t + \gamma_j \right\} } \\[2ex]
 & = & \frac{ \exp \{ \beta_{A0} + \beta_{A1}x_{At1} + \alpha_t + \delta_t \} \times \frac{1}{12} \sum_{j=1}^{12}\exp \left\{ \gamma_j \right\} }{ \exp \{ \beta_{N0} + \beta_{N1}x_{Nt1} + \alpha_t \} \times \frac{1}{12} \sum_{j=1}^{12}\exp \left\{ \gamma_j \right\} } \\[2ex]
  & = & \exp \{ \beta_{A0} + \beta_{A1}x_{At1} + \delta_t \} / \exp \{ \beta_{N0} + \beta_{N1}x_{Nt1} \} \\[2ex]
 & = & {RR}_0 \times \exp\{\beta_{A1}x_{At1} - \beta_{N1}x_{Nt1}\} \times \exp\{\delta_t\}
\end{array}
\end{equation}
where, for simplicity, we assume that there is only a single covariate of interest along with an intercept (i.e., ${\bf x}_{kt}^\top\bfbeta_k = \beta_{k0} + x_{kt1}\beta_{k1}$). From (\ref{approxRR}), the yearly risk ratio is approximated by three pieces: first, ${RR}_0 = \exp \left\{ \beta_{A0} - \beta_{N0} \right\}$, or the ``baseline'' risk ratio for the entire time interval; second, $\exp\{\beta_{A1}x_{At1} - \beta_{N1}x_{Nt1}\}$, a multiplicative scaling due to the covariates; and finally $\exp\{\delta_t\}$, a scaling for the risk ratio in a particular year. Note that the $\alpha_t$ term, which appears in both numerator and denominator and represents yearly deviations in event probabilities for both scenarios, cancels out. Furthermore, because the $\gamma_j$ are constant across scenario, the impact of seasonality also cancels out. Thus, we can interpret $\sigma^2 = \Var(\delta_t) \approx \Var(\log RR_t)$ as the effect of oceanic internal variability on the risk ratio or, alternatively, general variability in the $p_{At}$ above and beyond variability in the $p_{Nt}$. This single parameter represents the effect of prescribed ocean variability on the risk ratio after adjusting for a long-term trend by way of relevant covariate information.
}

%

\section{A diagnostic tool for event attribution sensitivity to oceanic variability} \label{diag1}

In addition to the sensitivity of the quantitative results, a natural question to ask in light of quantifying the effect of ocean variability is: do the qualitative results of an event attribution study change based on the effect of oceanic internal variability? 
Put another way, for the years considered in the analysis (1982-2013), for what proportion of years would the qualitative conclusion of an attribution analysis have changed based on the effect of ocean internal variability, assuming a present-day level of anthropogenic influence? 
In this framework, a ``change'' in an attribution analysis means concluding the presence of an anthropogenic effect (in one direction or another) instead of concluding the absence of an anthropogenic effect (or vice versa), and is relative to the event type and how strong the evidence must be to establish anthropogenic influence. 
It may also include changes in qualitative descriptions of the degree of anthropogenic influence, for instance ``moderate'' to ``small''.  The answer to this question may not be obvious from the numerical value of $\sigma$ alone, as it also depends on the proximity of $RR_{0}$ to the boundaries between qualitative categories.  Note that the statistical modeling approach allows us to impose a ``stationary'' climate (with present-day anthropogenic influence) by adjusting the probability estimates (and therefore the risk ratio estimates) by fixing each year to have the same covariate values, allowing comparison across each of the years. (For an illustration, see Section~\ref{global}.)

We formalize the question above by estimating $\pi$, the proportion of years for which the qualitative inference of a one-year, atmosphere-only attribution analysis remains the same over many different years (all arising from a stationary climate) in spite of internal variability. In other words, $\pi$ is the proportion of all years under consideration for which the risk ratio significantly exceeds (or does not exceed) a particular cutoff. Intuitively, large $\pi$ (near 1) indicates that the interval estimate of the risk ratio exceeds (or does not exceed) a threshold for most years; small $\pi$ (near 0) indicates that the interval estimate for the risk ratio exceeds (or does not exceed) the threshold for only a few years. For each region and event type, $\pi$ can be estimated using the MCMC samples (more details will be provided in Section \ref{global}).

We set our criterion as follows: if the interval estimate of $\pi$ for a particular region/event type and specified cutoff, say $(\pi_L, \pi_U)$, is entirely included in the set $[0, 0.05) \cup (0.95, 1]$, then the risk ratios are homogeneous (relative to the threshold) and the qualitative inference of an attribution analysis is insensitive to the specific year chosen for analysis. On the other hand, if the interval estimate for $\pi$ falls completely inside $[0.05, 0.95]$, then there is evidence of heterogeneity in both the risk ratio (relative to the threshold) and the corresponding qualitative statement about the risk. If the estimate of $\pi$ includes parts of both $[0, 0.05) \cup (0.95, 1]$ and $[0.05, 0.95]$, then while there is some evidence of heterogeneity, we are unable to make a definitive statement one way or the other. 

Therefore, we can use this tool to classify each region and event type into one of the following: in answer to the question ``Does the effect of ocean variability impact the consistency of attribution statements?'', either
\begin{enumerate}
\item Yes, the conclusions of an attribution study vary over time; i.e., $(\pi_L, \pi_U) \subset [0.05, 0.95]$.
\item Most likely yes, but the results are inconclusive; i.e., $(\pi_L, \pi_U) \not\subset [0.05, 0.95]$ but $(\pi_L, \pi_U) \cap [0.05, 0.95] \neq \{\varnothing\}$
\item No, the conclusions of an attribution study do not vary over time (relative to a threshold); i.e., either $(\pi_L, \pi_U) \subset [0, 0.05)$ or $(\pi_L, \pi_U) \subset (0.95,1]$.
\end{enumerate}
This strategy will be illustrated in Section \ref{global}.
 
\section{Case studies using climate model ensembles} \label{results}

The methods outlined in Section~\ref{methods} are illustrated with two case studies, both using the ensembles of the CAM5.1 atmosphere/land model from the C20C+~D\&A Project described in Section~\ref{app_fmwk}. The ensembles are used to estimate risk ratio for monthly (calendar month) hot, cold, and wet extremes;  dry extremes are omitted here because the zero-bound is encountered regularly for many regions on a monthly basis. 


\begin{figure}[!t]
\begin{center}
\vskip4ex
\includegraphics[trim={0 143 0 52mm}, clip, width = \textwidth]{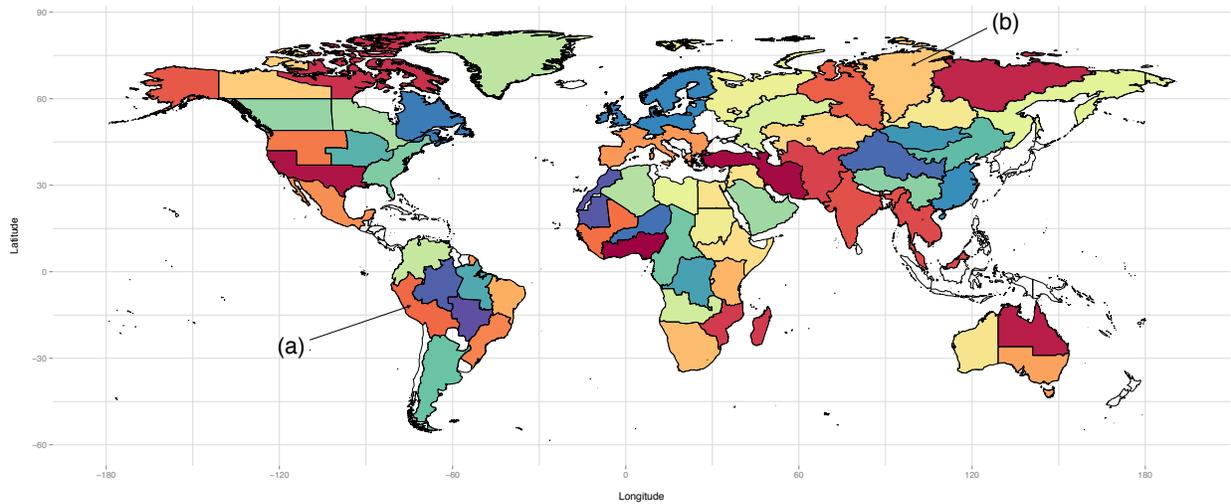}
\vskip1ex
\caption{Weather Risk Attribution Forecast (WRAF) regions ({\url{http://www.csag.uct.ac.za/~daithi/forecast}}). The two regions used as case studies are labelled as: (a) the southern Andean Community and (b) Krasnoyarsk.}
\label{WRAF}
\end{center}
\end{figure}

Temperature and precipitation are aggregated spatially into 58 geopolitical land-regions, which correspond to the regions used for the Weather Risk Attribution Forecast (WRAF; {\url{http://www.csag.uct.ac.za/~daithi/forecast}}), shown in Figure \ref{WRAF}. Within each region, temperature and precipitation are further aggregated by calendar month. For year $t$ and month $j$, we then have $n_t$ measurements, where $n_{t}$ is the number of simulations in year $t$. The one-in-ten year event for monthly data corresponds to the $100(1 - 0.1/12)^{\text{th}} = 99.17^{\text{th}}$ percentile for large extremes (hot, wet) and the $100(0.1/12)^{\text{th}} = 0.83^{\text{th}}$ percentile for small extremes (cold). In order to avoid an inappropriate bias from the more recent years with larger ensemble sizes, the empirical threshold percentiles were calculated from the 50-member ensemble that covers the entire time period of the study, from 1982-2013. For cold and wet events, the empirical percentiles were calculated from the ALL simulations; for hot events, the empirical percentiles were calculated from the NAT simulations (in order to ensure sufficient sampling in the ALL simulations, as one in ten year hot events in ALL typically do not occur in NAT). Our interest here is in the probability (and risk ratio) of extreme monthly events, irrespective of when they occur during the year; hence we average over the monthly probabilities to obtain the yearly probabilities as defined in Section \ref{rrm}.  

\begin{figure}[!t]
\begin{center}
\includegraphics[width=\textwidth]{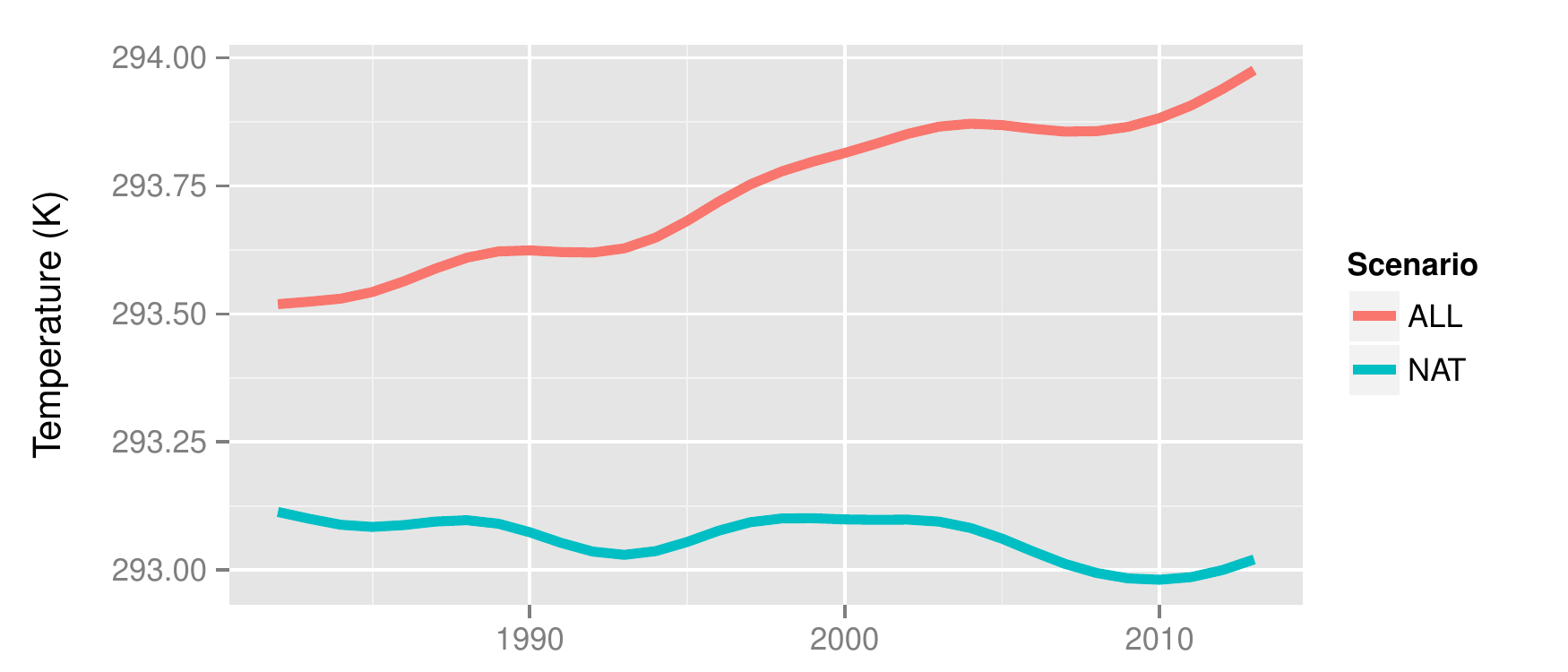}
\caption{Global mean temperature measurements between 50$^\circ$S and 50$^\circ$N, calculated as an land-ocean ensemble average from the 50 members that covered the entire time period from 1980 to 2013, using the low-pass filter described in Appendix 3.A of \cite{IPCC2007}.}
\label{temps}
\end{center}
\end{figure}

The case studies explore the risk ratio from year to year, and consider two WRAF regions (labelled in Figure~\ref{WRAF}): (a) the southern Andean Community (comprising Peru and Bolivia) and (b) Krasnoyarsk (a federal subject in central Russia). As will be illustrated, these two regions were chosen as case studies because they exhibit very different behavior in monthly extremes and their relation to sea surface conditions. The empirical probabilities of each event type for the southern Andean Community and Krasnoyarsk are provided in Appendix~\ref{suppFig}. 

Recall that the $\delta_t$ terms in (\ref{full_model}) capture variation above and beyond an overall trend in the $RR_t$. As a way to describe the overall trend in the extreme event probabilities, a scenario-specific global mean temperature covariate was used. More specifically, the covariate used is the average surface air temperature over land and ocean between 50$^\circ$S and 50$^\circ$N, calculated as an ensemble average from the 50 members that covered the entire time period from 1982 to 2013 (the average was smoothed using the the low-pass filter described in Appendix 3.A of \citealp{IPCC2007}). The scenario-specific values are plotted in Figure \ref{temps}.  The use of actual values, rather than say a linear trend fit, means that we are making no assumptions of the temporal shape of the climatic response to external forcings;  the latitude range prevents strong influence from small variations in the sea ice edge.

\begin{figure}[!t]
\begin{center}
\includegraphics[width=\textwidth]{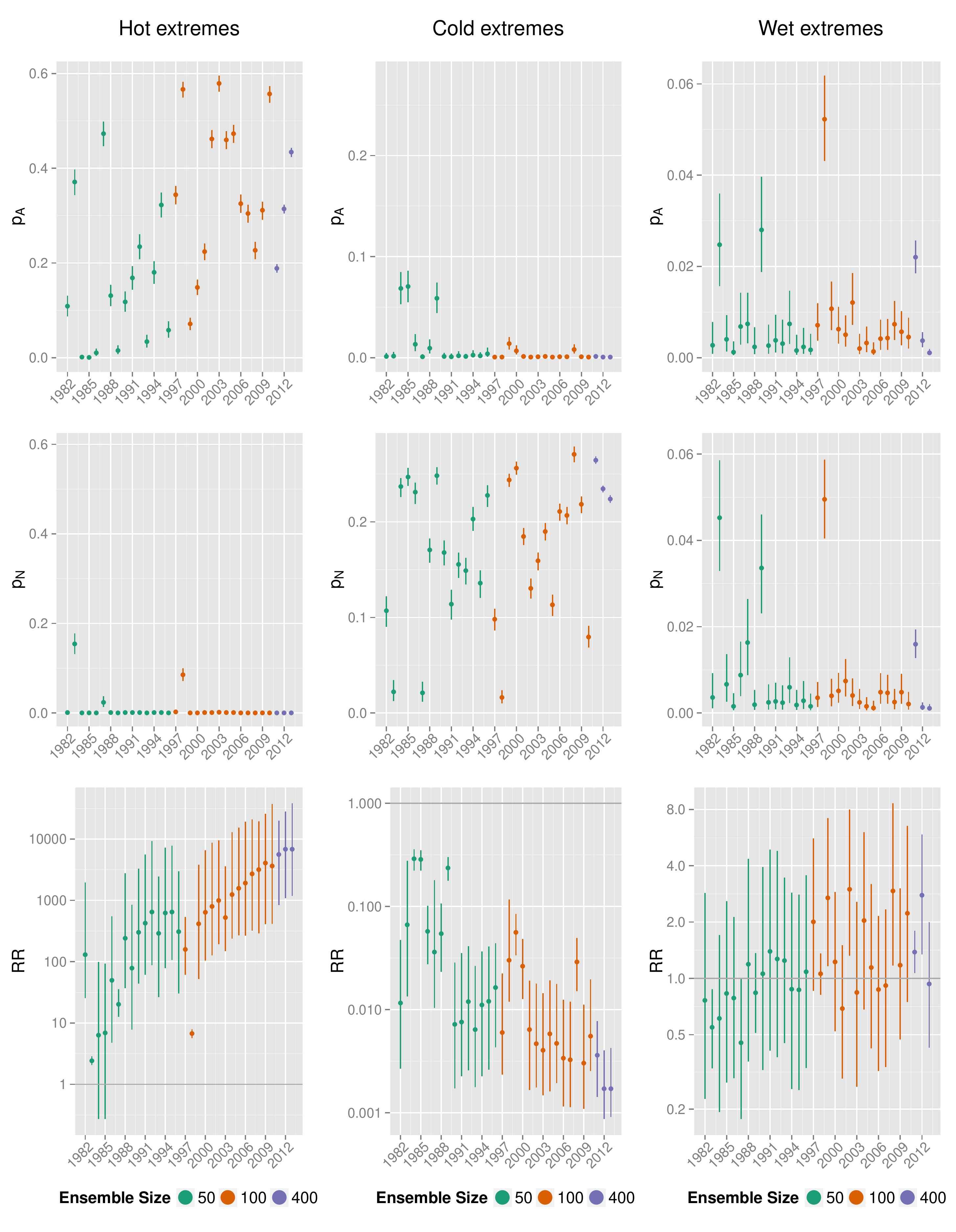}
\caption{Probabilities and risk ratio estimates for extremes in the southern Andean Community.}
\label{SA_estimates}
\end{center}
\end{figure}

\begin{figure}[!h]
\begin{center}
\includegraphics[width=\textwidth]{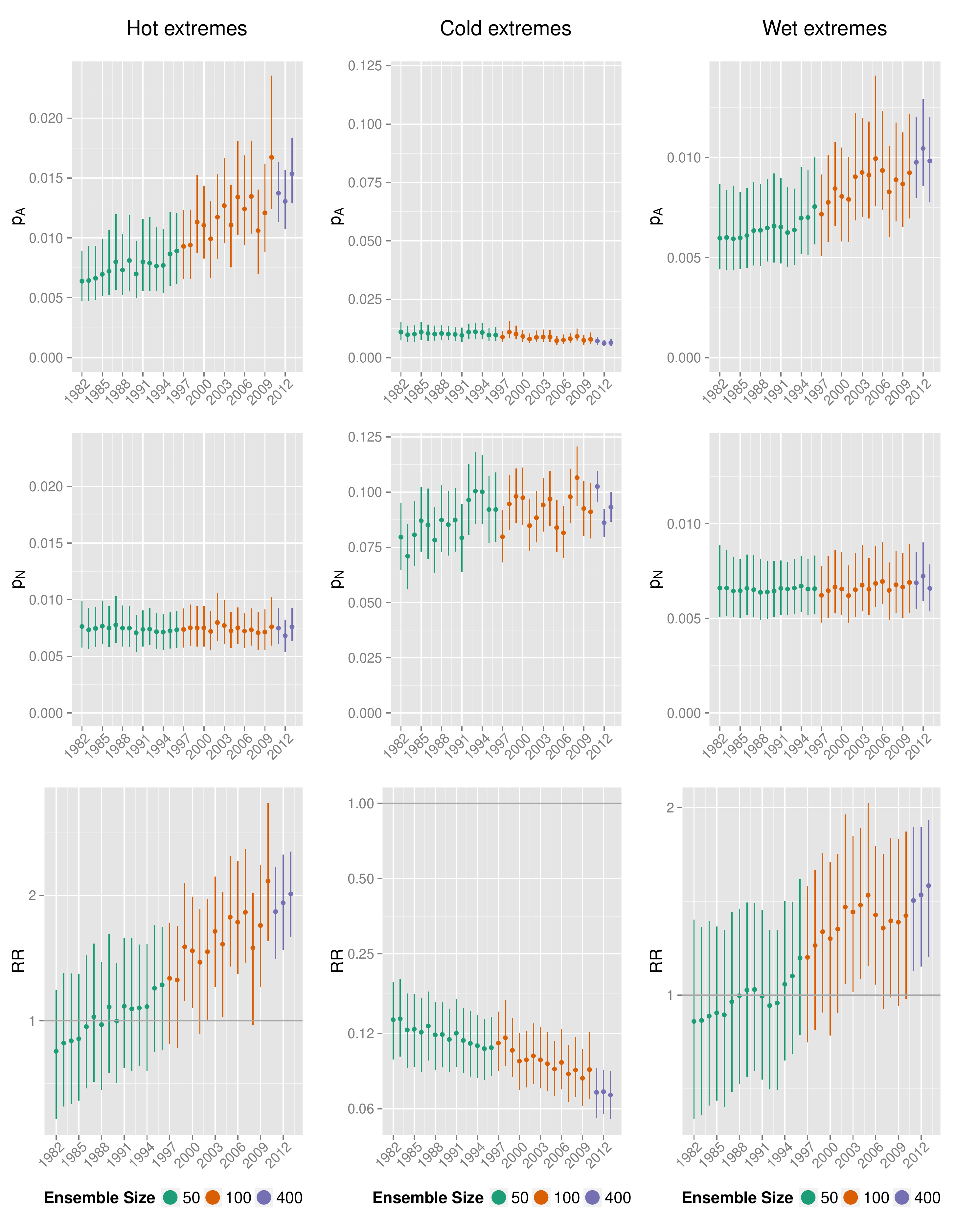}
\caption{Probabilities and risk ratio estimates for extremes in Krasnoyarsk, Russia.}
\label{KR_estimates}
\end{center}
\end{figure}

\subsection{Case Study 1: southern Andean Community}

\begin{figure}[!t]
\begin{center}
\includegraphics[width=\textwidth]{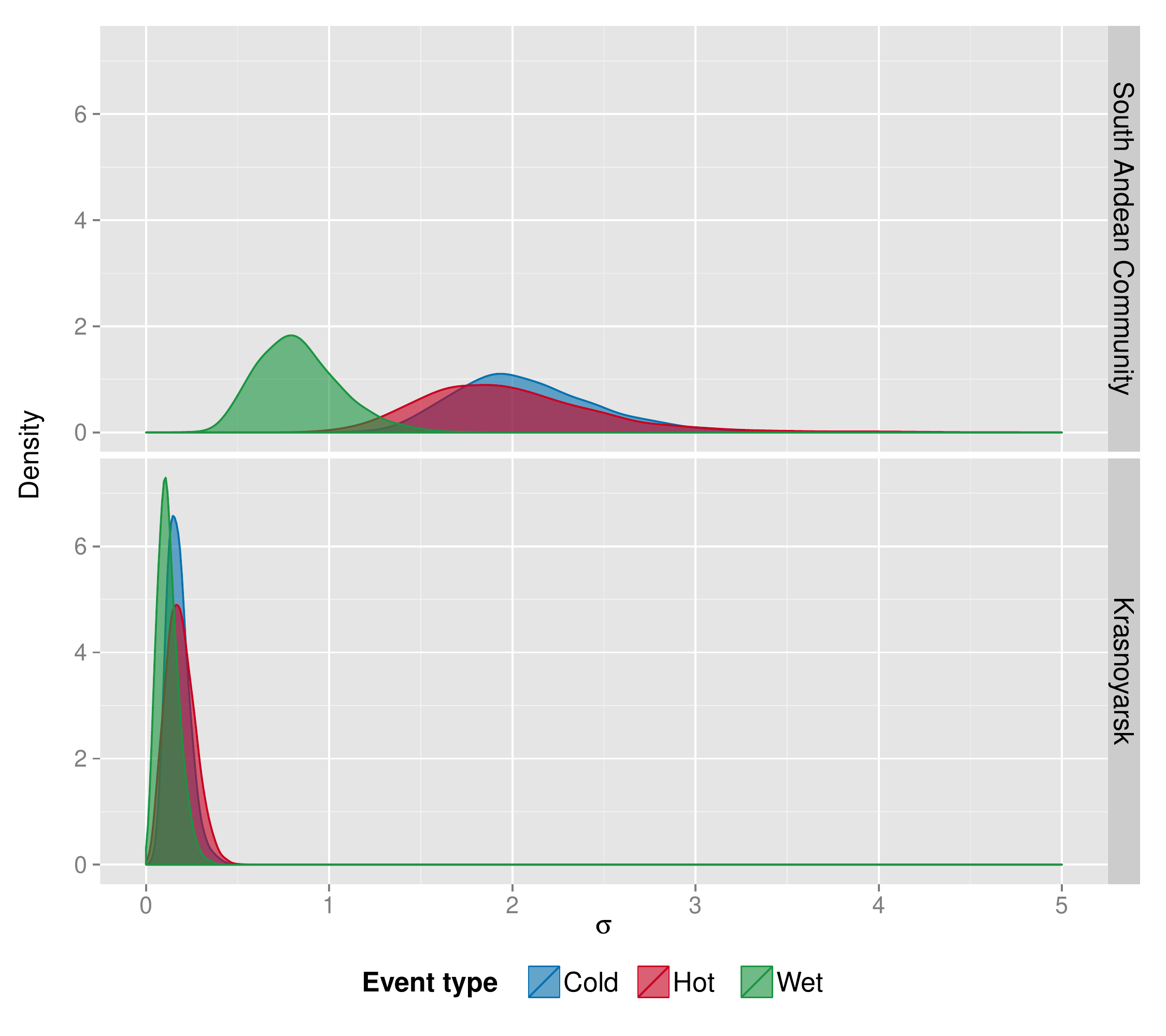}
\caption{Posterior distributions for $\sigma$, the effect of ocean variability on log risk ratio, for each event type in the southern Andean Community (top) and Krasnoyarsk (bottom).}
\label{both_sds}
\end{center}
\end{figure}

The temporal estimates of the event probabilities and risk ratio for each event type over the southern Andean Community are shown in Figure~\ref{SA_estimates}. 
For hot extremes note that high temperature events are extremely common in the ALL forcings scenario (the $p_{At}$ are as large as 0.6) but very unlikely in the NAT scenario, yielding median risk ratio estimates and also lower bounds on those estimates that are extremely large (with $p_N\approx0$, $RR \approx \infty$). 
Even though they occur on average 0.83\% of the time (by definition), hot events for NAT are essentially only possible during large El Ni\~no years (e.g., 1983 and 1998), but even then the risk ratio estimates are significantly larger than 1. Even after adjusting for the long term trend imposed by the changes in global temperature, the posterior distribution for $\sigma$ for hot events in Figure \ref{both_sds} shows that ocean variability has a very large impact on the log risk ratio, the largest for any event type in this region. This is not surprising, as the southern Andean Community is known to be strongly affected by ocean conditions in the eastern tropical Pacific.

In terms of the probability and risk ratio estimates, Figure \ref{SA_estimates} shows that cold extremes have essentially the opposite behavior of hot extremes for this region: extremely cold events are common in the NAT scenario and very uncommon (on average 0.83\% by definition) in the ALL scenario. As a result, the risk ratio estimates are extremely small and significantly less than one for all years, particularly for the most recent years. From Figure \ref{both_sds} we see that there is still a large effect of the ocean variability on log risk for cold events, albeit somewhat less so than for hot events.

The results for precipitation are perhaps the most interesting for the southern Andean Community. For wet extremes, the event probabilities in each scenario are much more comparable, leading to risk ratio estimates much closer to 1. For the ALL scenario, it is quite easy to pick out the biggest El Ni\~no events, as these years yield the largest event probabilities (1983, 1989, 1998, and 2011). 
However, note that these years also correspond to large probabilities in the NAT scenario, with the result that the median risk ratio estimates are in fact comparable to neighboring years. Because the ratio of the uncertainty in both $p_A$ and $p_N$ to their median value is smaller for higher probabilities (i.e., El Ni\~no years), the corresponding uncertainty in the risk ratio estimates is also smaller.
Finally, while the effect of ocean variability on the log risk ratio is smaller for wet events than hot or cold (Figure \ref{both_sds}), the effect is decidedly non-zero; this is visible in the estimates of risk ratio in Figure \ref{SA_estimates} with respect to a number of non-overlapping interval estimates (e.g., 1983 vs. 2011).


\subsection{Case Study 2: Krasnoyarsk, Russia}

Similar plots of the estimated probabilities, estimated risk ratio, and posteriors for $\sigma$ in Krasnoyarsk are shown in Figures \ref{KR_estimates} and \ref{both_sds}.

For each event type, the effect of ocean variability is greatly reduced for Krasnoyarsk relative to the southern Andean community, which is not at all surprising given the high latitude continental nature of the region. Hence, individual El Ni\~no/La Ni\~na years do not stand out in the same way as for the southern Andean community. This is in spite of the estimates of the risk ratio having visible trends over time, the implication being that this trend is almost completely described by the long-term trend in global mean temperature (i.e., the covariate). There is a large difference between scenarios for the probabilities of cold extremes for Krasnoyarsk, leading to risk estimates that are significantly less than 1. The fact that this is the case even for the 1980s may be related to the effect of Soviet-era aerosol pollution on winter cold spells (but which is outside the consideration of this paper).

\section{Global results summarizing the effect of ocean variability} \label{global}

To summarize the effect of ocean variability more broadly for all of the WRAF regions, we provide one-dimensional ``maps'' (sorted by central latitude) of the posterior distributions of $\sigma$ for each event type, which show the median estimate as well as the 95\% probability interval (called a ``credible interval'' in a Bayesian framework, as opposed to confidence interval); see Figures \ref{hotIV}, \ref{coldIV}, and \ref{wetIV}. 

However, before we describe the color schemes, we first discuss how the shadings are determined. As mentioned in Section \ref{diag1}, for each region and event type, we can determine whether or not the effect of ocean variability causes different conclusions to be drawn from attribution studies in different years. In order to do this, we need to adjust the probability and risk ratio estimates such that the different years behave as if they occurred under a stationary climate, which is done as follows. From (\ref{riskratio_calc}), with a single covariate (global mean temperature), the (unadjusted) risk ratio for a particular year $t$ is calculated as
\begin{equation*} \label{riskratio_calc2} 
RR_t =  \frac{ \frac{1}{12} \sum_{j=1}^{12} \logit^{-1}(\beta_{A0} + \beta_{A1} x_{At} + \alpha_t + \delta_t + \gamma_j)}{ \frac{1}{12} \sum_{j=1}^{12} \logit^{-1}(\beta_{N0} + \beta_{N1} x_{Nt}  + \alpha_t + \gamma_j)}.
\end{equation*}
To transform this estimate to behave as though it arose from a stationary climate, we can substitute common covariate values $x_A^*$ and $x_N^*$ for the year-specific values $x_{At}$ and $x_{Nt}$, while maintaining the year-specific effects $\alpha_t$ and $\delta_t$ to allow for year-to-year differences within the stationary climate state. That is, we can instead use the adjusted risk ratio $\widetilde{RR}_t$, where
\begin{equation} \label{riskratio_adj} 
\widetilde{RR}_t = \frac{ \frac{1}{12} \sum_{j=1}^{12} \logit^{-1}(\beta_{A0} + \beta_{A1} x_A^* + \alpha_t + \delta_t + \gamma_j)}{ \frac{1}{12} \sum_{j=1}^{12} \logit^{-1}(\beta_{N0} + \beta_{N1} x_N^*  + \alpha_t + \gamma_j)}
\end{equation}
for $t=1982, \dots, 2013$ to get a sense of what the risk ratio would look like over the time period had there been no changes to the climate state. Specifically, we chose to set $x_k^*$ equal to the average global mean temperature in scenario $k\in \{A,N\}$ from the five most recent years (2009-2013).

\begin{figure}[!t]
\begin{center}
\includegraphics[width=\textwidth]{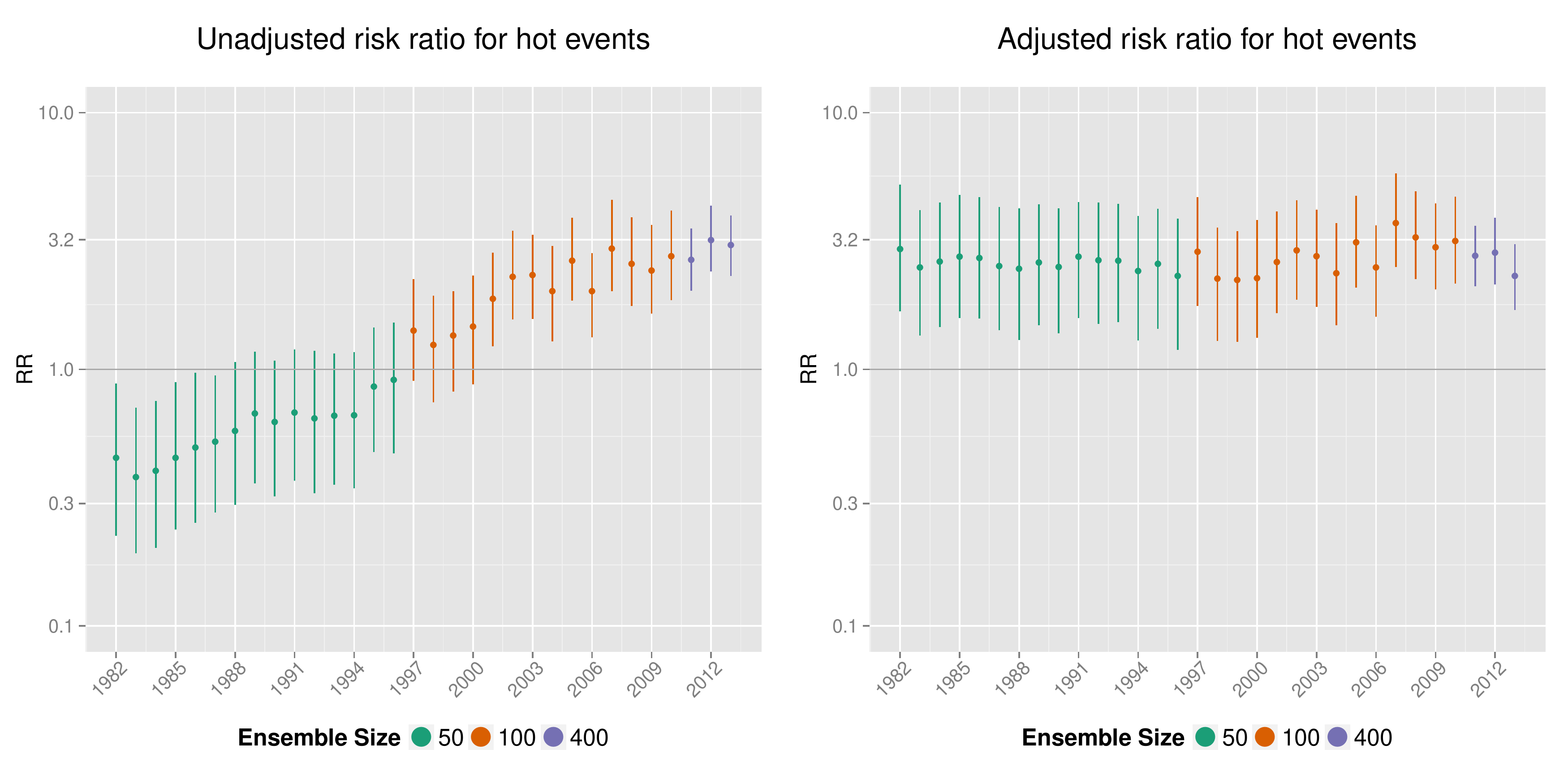}
\caption{Unadjusted ($RR_t$; left) and adjusted ($\widetilde{RR}_t$; right) risk ratio for hot events in the northern European Economic Area (EEA), shown as the blue region in northern Europe in Figure \ref{WRAF}. The adjusted risk is calculated using a fixed covariate value (global mean temperature) for all years.}
\label{hot_adjRR}
\end{center}
\end{figure}

As an illustration of how this changes the risk ratio estimates, consider the plot in Figure \ref{hot_adjRR}, which shows the raw (unadjusted) risk ratio estimates and the adjusted risk ratio estimates for hot events in the northern European Economic Area (EEA), shown as the blue region in northern Europe in Figure \ref{WRAF}. This region is unusual in that it has experienced strong anthropogenic influence (via land cover change and production of various types of aerosols) for a much longer time, with the balance of these effects resulting in a summer cooling effect prior to the 1990s in the CAM5.1 simulations.  This late crossover exists in climate models generally, but with different timings, and means that risk ratio calculations for European events can be sensitive to the definition of the counterfactual ``natural'' climate.

Using the adjusted risk ratios, we are now ready to estimate the $\pi$ probabilities for each region and event type. As mentioned in Section~\ref{diag1}, $\pi$ is estimated using each of the MCMC joint posterior samples. Each MCMC sample yields a point-estimate time series of the estimated risk; aggregated over all posterior samples, these time series define the 95\% credible intervals shown in, e.g., Figure \ref{hot_adjRR}. For each sample, we then calculate the proportion of years (out of the total $T=32$) for which the risk ratio exceeds (or does not exceed) a particular cutoff. Then, aggregating over all of the MCMC samples, we can obtain a posterior distribution on $\pi$ for each region and event type. Once we have the posteriors, we are prepared to classify each region as outlined in Section~\ref{diag1}.

\begin{figure}[!t]
\begin{center}
\includegraphics[width=\textwidth]{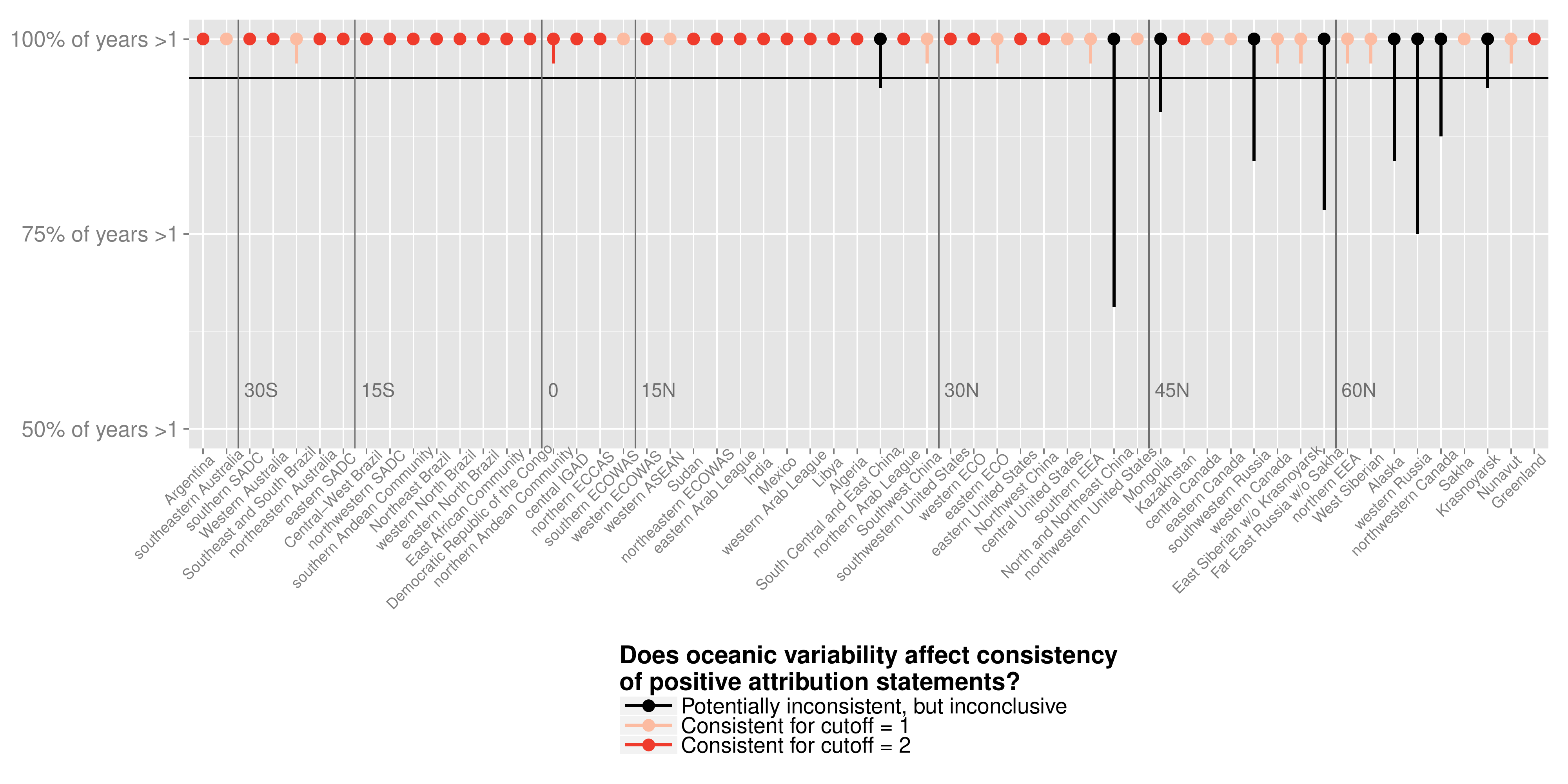}
\caption{95\% interval estimates of $\pi$ for a exceedance threshold of 1 for hot events, with the horizontal black line denoting the 0.95 cutoff for $\pi$. The intervals represent an estimate of the proportion of years for which the risk ratio is greater than 1, assuming a stationary climate (i.e., using $\widetilde{RR}_t$). Black intervals represent regions for which oceanic variability potentially impacts the consistency of attribution statements over time. The varying shades of red, on the other hand, are used to denote regions for which oceanic variability does \textit{not} affect consistency of attribution statements, for several different thresholds (i.e., $RR>1$, $RR>2$, and $RR>10$).}
\label{hot_EP}
\end{center}
\end{figure}

\begin{figure}[!t]
\begin{center}
\includegraphics[width=\textwidth]{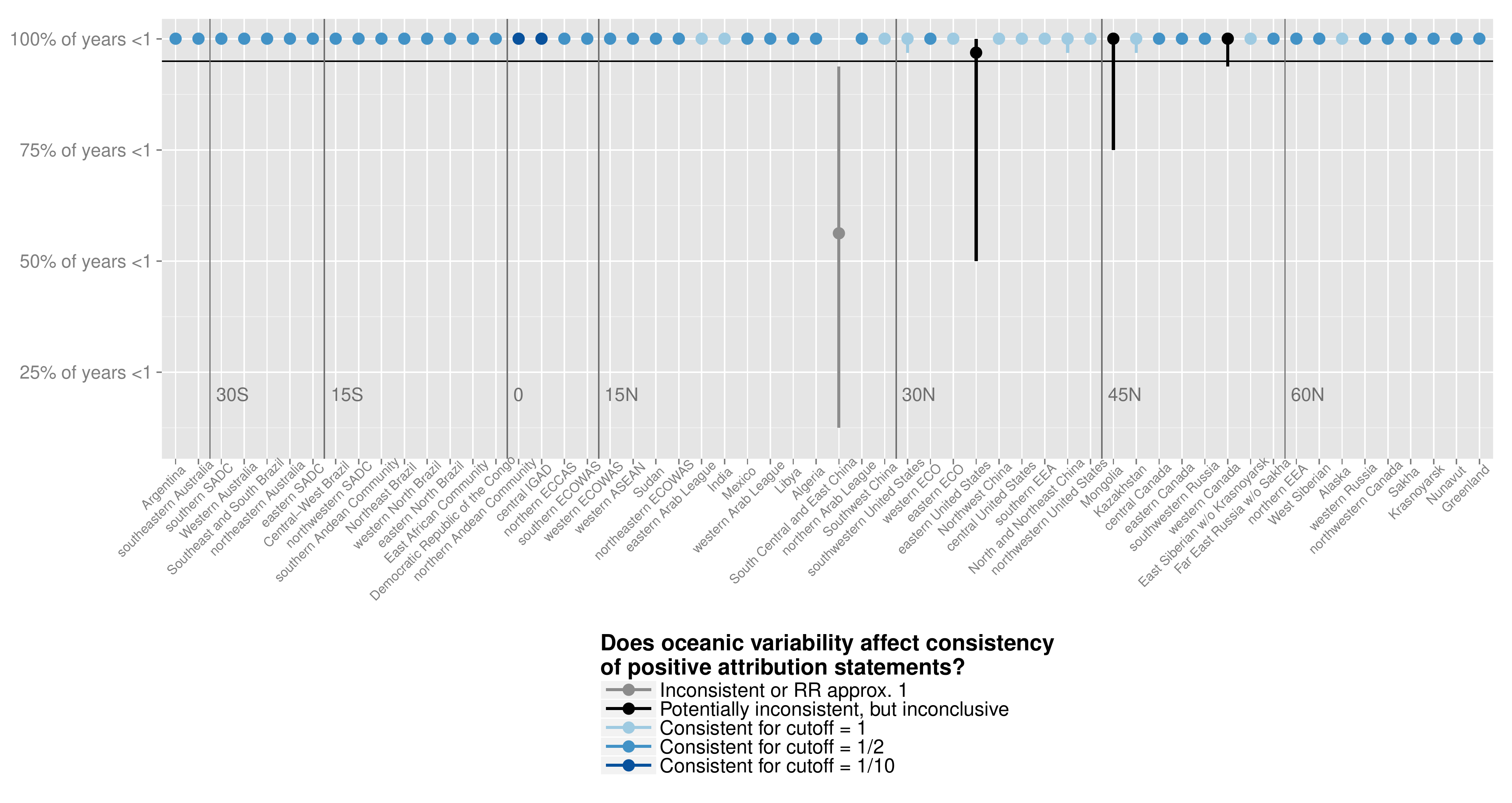}
\caption{95\% interval estimates of $\pi$ for \textit{not} exceeding a threshold of 1 for cold events, with the horizontal black line denoting the 0.95 cutoff for $\pi$. The intervals represent an estimate of the proportion of years for which the risk ratio is less than 1, assuming a stationary climate (i.e., using $\widetilde{RR}_t$). Black intervals again represent regions for which oceanic variability potentially impacts the consistency of attribution statements over time. Alternatively, the varying shades of blue are used to denote regions for which oceanic variability does \textit{not} affect consistency of attribution statements, for several different thresholds (i.e., $RR<1$, $RR<1/2$, and $RR<1/10$).}
\label{cold_EP}
\end{center}
\end{figure}

\begin{figure}[!t]
\begin{center}
\includegraphics[width=\textwidth]{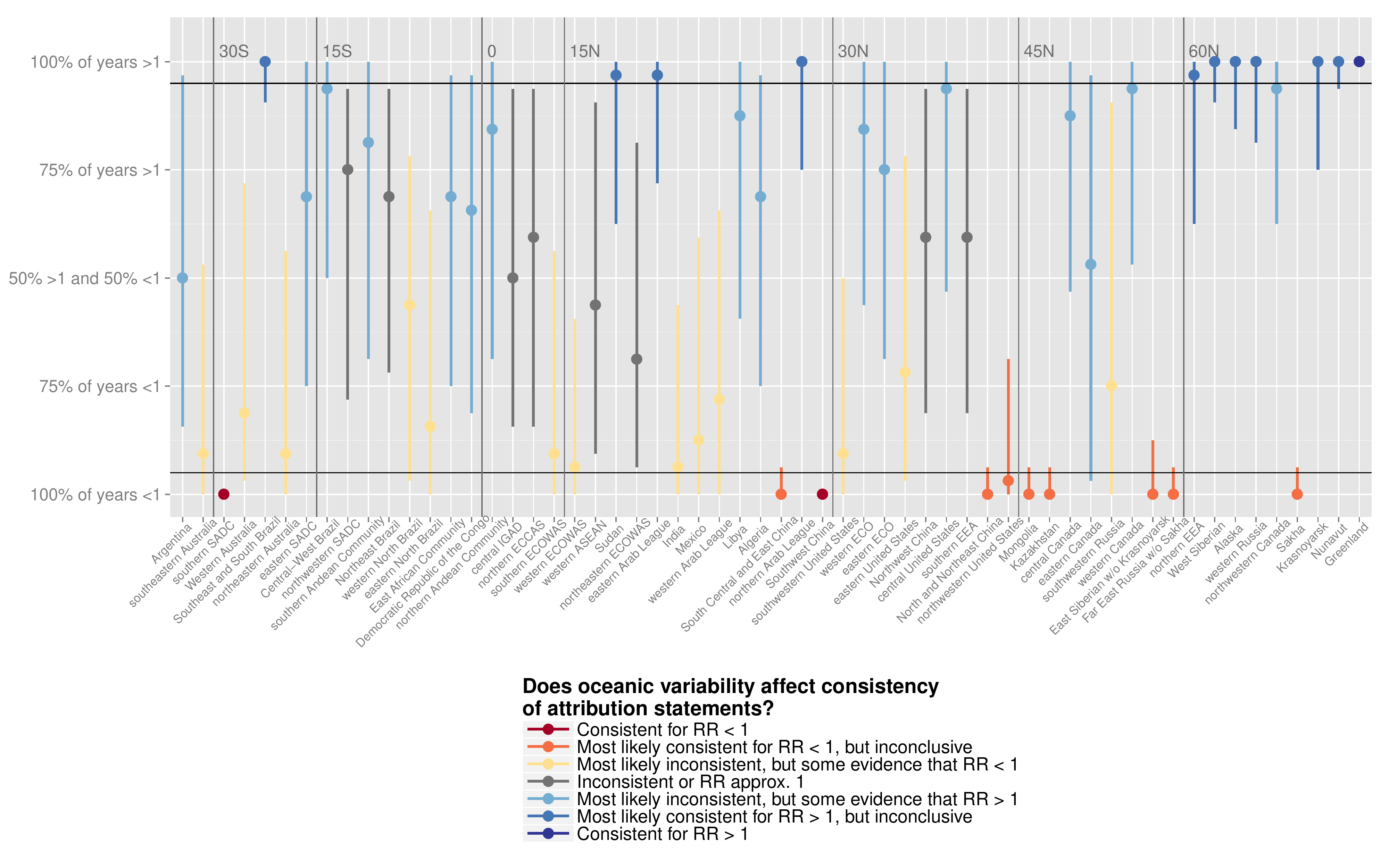}
\caption{95\% interval estimates of $\pi$ for a exceedance threshold of 1 for wet events. The intervals represent an estimate of the proportion of years for which the risk ratio is greater than 1, assuming a stationary climate (i.e., using $\widetilde{RR}_t$). Gray intervals now represent regions for which oceanic variability \textit{does} impact the consistency of attribution statements over time, or, alternatively, the risk ratio is close enough to unity that variations above and below 1 are insignificant in light of sampling variability. Blue (yellow/red) shadings represent regions for which there is a varying degree of evidence for a consistent increase (decrease) in the probability of an extremely wet month, with darker colors corresponding to stronger evidence.}
\label{wet_EP}
\end{center}
\end{figure}

\begin{figure}[!t]
\begin{center}
\includegraphics[width=\textwidth]{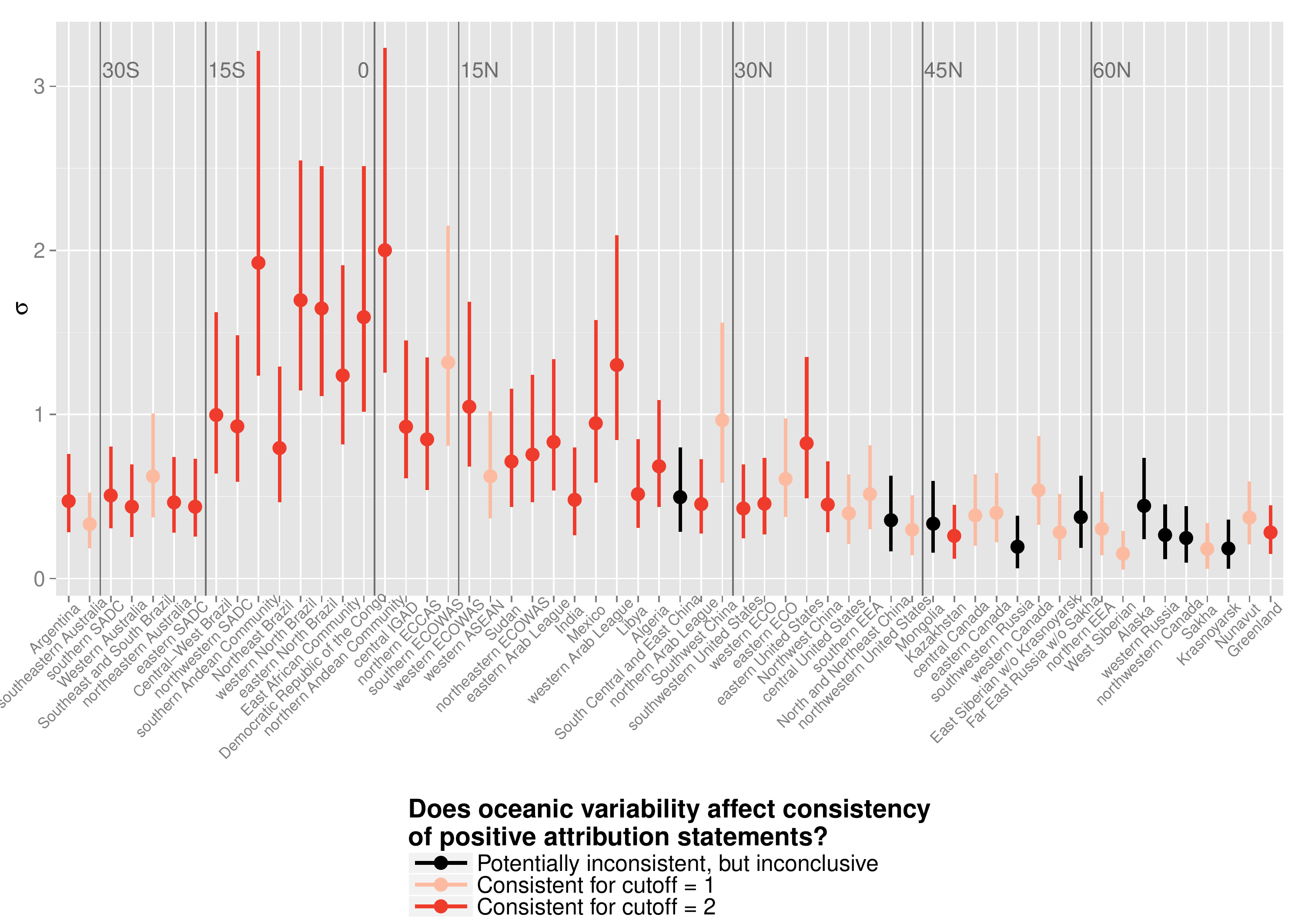}
\caption{95\% interval estimates of $\sigma$, the effect of ocean variability, for hot events across WRAF regions. The colors correspond to those in Figure \ref{hot_EP}.}
\label{hotIV}
\end{center}
\end{figure}

\begin{figure}[!h]
\begin{center}
\includegraphics[width=\textwidth]{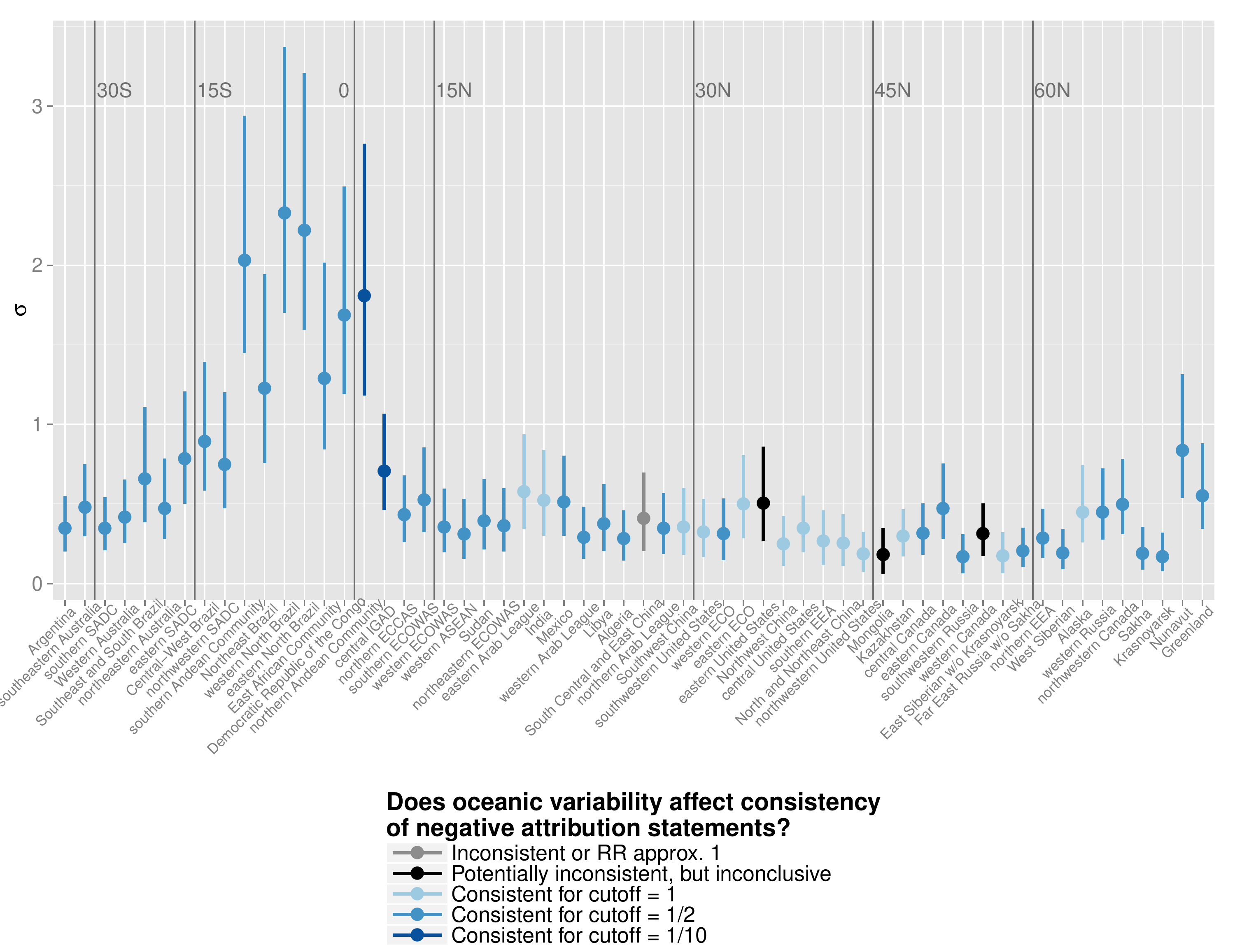}
\caption{95\% interval estimates of $\sigma$, the effect of ocean variability, for cold events across WRAF regions. The colors correspond to those in Figure \ref{cold_EP}.}
\label{coldIV}
\end{center}
\end{figure}

\begin{figure}[!h]
\begin{center}
\includegraphics[width=\textwidth]{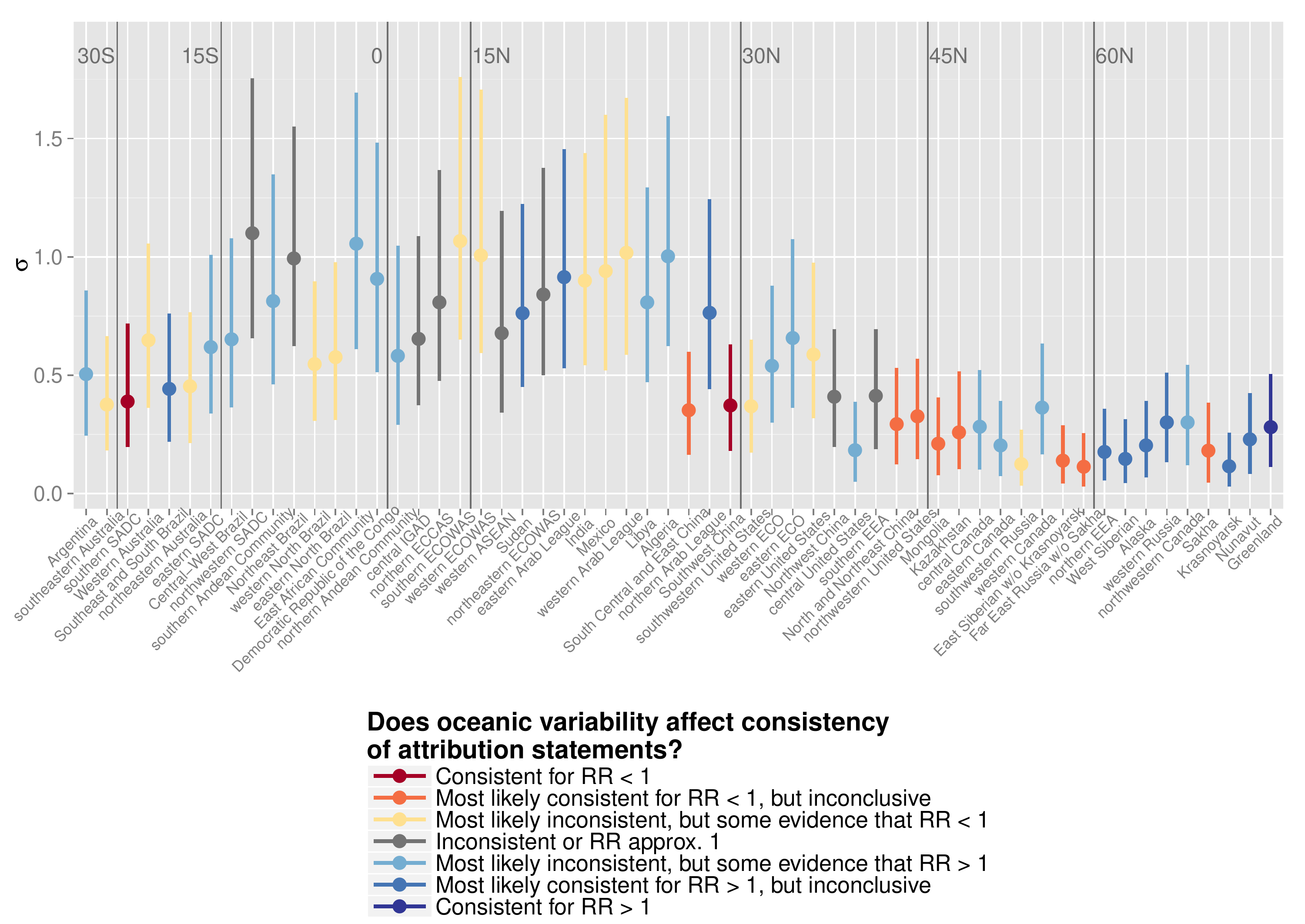}
\caption{95\% interval estimates of $\sigma$, the effect of ocean variability, for wet events across WRAF regions. The colors correspond to those in Figure \ref{wet_EP}.}
\label{wetIV}
\end{center}
\end{figure}

The interval estimates of $\pi$ across all regions for each event type are shown in Figures \ref{hot_EP}, \ref{cold_EP}, and \ref{wet_EP} and are shaded according to the classification described in Section \ref{diag1}. Then, using the same shading, estimates of the effect of ocean variability on the results of a one-year, atmosphere-model-only attribution study (i.e., $\sigma$) are shown in Figures~\ref{hotIV}, \ref{coldIV}, and \ref{wetIV}. The shaded plots of $\sigma$ highlight the somewhat counterintuitive relationship between the magnitude of the ocean variability's impact and the corresponding impact on qualitative statements over time.
The effect of anthropogenic emissions is now strong enough that the conclusions that $RR>1$ for hot events, and $RR<1$ for cold events, are rarely challenged by considering the effect of ocean variability;  even the $RR>2$ or $RR<\frac{1}{2}$ conclusions still hold for most regions.  What is perhaps surprising is that the regions most affected lie in the mid- and high-latitudes, where the All-Hist probabilities ($p_{A}$) are largely unaffected by SSTs (as reflected in the low skill of seasonal weather forecasts over these regions) and the $\sigma$ values are thus quite low.  In contrast, the regions with the most robust conclusions of an anthropogenic influence, with $RR>10$ or $RR<\frac{1}{10}$ being robust in many cases, lie in the tropics, where the effect of SSTs on $p_{A}$ is high (as reflected in the relatively high skill of seasonal forecasts) and the $\sigma$ values are much higher.  In other words, conclusions based on classification into the qualitative influence of anthropogenic emissions are strongly anticorrelated to conclusions based on numerical sensitivity (as measured by $\sigma$) in terms of their sensitivity to the effect of interannual ocean variability.

Are these results unexpected?  The three case studies illustrated in Section~\ref{results} and Figure~\ref{hot_adjRR} indicate why this should not be a surprise.  Temperatures in tropical areas are controlled by radiative processes, and thus directly by imposed radiative boundary conditions, whereas temperatures in extratropical areas are also heavily influenced by dynamical processes.  This means that interannual variability is lower over tropical land, and thus that for the same amount of mean warming the risk ratio is much higher there than over the extratropics, all other things being equal.  For instance, the hot and cold risk ratios for the southern Andean Community are well beyond 100 and $\frac{1}{100}$ respectively, in recent years, whereas they are within 10 and $\frac{1}{10}$ for Krasnoyarsk and the northern European Economic Area.  Hence, tropical regions have much more leeway to accommodate the moderately larger (as in a factor of three rather than several orders of magnitude) uncertainty described by $\sigma$, whereas many extratropical regions do not.  This result may not hold if the hot categories were defined by, for instance, whether $RR>10^{4}$, but we argue that such a definition would be pedantic in the light of any practical interpretation of the role of anthropogenic emissions.

The comparison of the quantitative and qualitative conclusions for wet months differs somewhat from the comparisons for hot and cold months.  In this case there is no discernible geographic pattern in the influence of SST variability, at least in part resulting from the fact that the estimated risk ratios are close to -- and consistent with -- unity in the first place.  However, there is also a similarity in that $\sigma$ values are higher in the tropics;  while there is no anticorrelation in this case, the $\sigma$ values themselves are poor indicators of the categories. Given the lack of relationship between latitude and the color categories in Figures \ref{wet_EP} and \ref{wetIV}, we provide a map of the regions shaded by their color group (see Figure \ref{wetMap}).

\section{Discussion} \label{discussion}

This study was conducted to understand, both quantitatively and qualitatively, how many recent and current studies assessing the role of anthropogenic emissions on individual extreme atmospheric events may be biased because their adoption of an experimental design that conditions on the observed ocean state.  While quantitative estimates for events in the tropics are found, as expected \citep{OttoFEL_BoydE_etalii_2015}, to be more sensitive to the conditioning on the ocean state, in fact for hot and cold events the qualitative effect on conclusions is larger in the extratropics.  While tests were performed only for events over a certain set of regions, all approximately 2~million~km$^{2}$ in size, and covering a calendar month, results should be broadly transferrable to events covering overlapping regions of varying sizes and lasting for different durations in most cases \citep{AngelilO_StoneDA_etalii_2014}, with the possible exception of small-scale, short-duration wet events \citep{AngelilO_StoneDA_PallP_2014}.

In Section~\ref{introduction} we noted that the atmospheric modeling approach to diagnosing the role of anthropogenic emissions in extreme weather, as introduced by \citet{PallP_AinaT_etalii_2011}, rests on three major assumptions in addition to uncertainties common to other approaches:  that results are invariant to ocean state;  that there is no change in ocean variability;  and that coupled atmosphere-ocean interactions are unimportant.  At first glance, these three assumptions may be expected to be more valid for the extratropics than the tropics.  Concerning the third assumption, winds can more easily alter the SSTs in the tropics due to the shallower thermocline there, meaning that the third assumption should be more questionable in the tropics.  It is not clear however that this assumption introduces any appreciable bias into estimates of the role of anthropogenic emissions.  Concerning the second assumption, year-to-year ocean variability affects the tropical atmosphere more directly -- as evidenced by the higher skill of tropical seasonal forecasts -- and thus any change in ocean variability driven by anthropogenic emissions could dominate the effect of anthropogenic emissions on tropical terrestrial extremes.  For instance, given that unusually wet months only occur in Peru during El~Ni\~no events (Figure~\ref{SA_estimates}), a marginal increase or decrease in the frequency or intensity of El~Ni\~nos due to anthropogenic emissions would have a direct influence on the probability of a wet month.

The greater role of the SSTs in the tropics might intuitively be expected to make the first assumption more questionable in the tropics.  If the probability of a cold extreme is strongly influenced by the occurrence of an El~Ni\~no event, then it would seem reasonable to hypothesize that the calculation of the risk ratio from an atmospheric modeling experiment, which is the ratio of two probabilities of hot extremes, might also be affected.  Indeed, this appears to be the case for the southern Andean community (Figure~\ref{SA_estimates}) but the effect is not large.  More generally, such an effect only appears to be $\frac{3}{2}$ to 4 times larger in the tropics than in the extratropics.  In contrast, the effect of lower year-to-year variability in temperature in the tropics has a much larger impact on the risk ratio.  As a consequence, while the validity of the first assumption of invariance to ocean variability may be weaker in the tropics, qualitative conclusions of the degree of anthropogenic influence are generally far more sensitive in the extratropics.  This comparison of between the tropics and extratropics is summarised in Table~\ref{TAB_SUMMARY}.

\begin{table}[!t]
  \caption{Comparison between the importance of the three major assumptions underlying the atmospheric modelling approach to diagnosing the role of anthropogenic emissions in extreme weather, as distinguished between the tropics and extratropics.}
\begin{tabular}{|p{0.12\linewidth}|p{0.38\linewidth}|p{0.38\linewidth}|}
    \hline
    \textbf{Event type} & \textbf{Tropics} & \textbf{Extratropics}\\
    \hline \hline
    \multicolumn{3}{|l|}{\sl 1. Results are invariant to ocean state} \\ \hline\hline
    Hot and cold months & Technically invalid, but not relevant in practice & Validity unclear, but can be quite important in practice \\ \hline
    Wet months & Technically invalid, could be important factor & Validity unclear, may be important in practice \\
    \hline \hline
    \multicolumn{3}{|l|}{\sl 2. No change in ocean variability} \\ \hline\hline
    Hot and cold months & Validity questionable due to some control from ocean variability, possibly dominant factor & Weak control from ocean variability, so likely valid \\  \hline
    Wet months & Not valid due to strong control from ocean variability, possibly dominant factor & Weak control from ocean variability, so likely valid \\
    \hline \hline
    \multicolumn{3}{|l|}{\sl 3. Coupled atmosphere-ocean interactions are unimportant} \\ \hline\hline
    Hot, cold, and wet months & Stronger interactions, importance not clear, perhaps invalid for some events (e.g. tropical cyclones) & Weak interactions, so likely valid \\
    \hline
  \end{tabular}
  \label{TAB_SUMMARY}
\end{table}

In order to assess the effect of using atmosphere-only climate models on attribution statements for extremes, 
using a Bayesian hierarchical statistical approach we developed a corresponding decision model for estimating the probability of exceedance of a threshold given a large set of related ensembles of simulations, as well as for estimating the risk ratio estimated from pairs of such probabilities.  This model should have broader application beyond the analysis performed in this paper.  For instance, if climate model simulations are available for multiple years, as is the case in the C20C+~D\&A Project, then more accurate estimates of the posterior probabilities and the risk ratio are possible by incorporating data from multiple years.  The use of a covariate not only provides the possibility for formal detection of long-term trends in extremes, but also for detecting the influence of other factors such as volcanic eruptions and El~Ni\~no events. In addition to using data from multiple years and including covariates, it may also be possible to borrow information from across multiple estimates of the ocean warming attributable to anthropogenic emissions and from across various atmospheric models;  however, in practice the selection of the prior distribution would be less obvious in these cases than it is across multiple years.

Finally, a more direct application of the results presented in this paper could be in the assessment of the robustness of calculated attribution conclusions to the effect of ocean variability. In other words, the estimates provided in this paper of the effect of oceanic variability could potentially be combined with estimates of the risk ratio made by future single-year, atmosphere-only attribution studies to make a statement about the ``true'' risk ratio, that is, the risk ratio informed by all possible years. 
The spatio-temporal characteristics of observed events are unlikely to exactly coincide with the monthly-region definitions used in this paper.  However, event attribution results appear to be transferrable in a predictable way for both the spatial and temporal definitions for hot and cold events and to some degree in the spatial definition for wet events \citep{AngelilO_StoneDA_etalii_2014}, although perhaps not at the local scale for wet events \citep{AngelilO_StoneDA_PallP_2014}.

As a concrete example, suppose a hypothetical research team conducts a one-year, atmosphere-only attribution analysis that involves estimating the risk ratio for a certain event type in a certain region \citep[for example in ][]{PallP_AinaT_etalii_2011}. Furthermore, suppose that their new region can be associated with one of the regions used in our analysis (see Section~\ref{results}), and that the study concerns one of the atmospheric variables used in Section~\ref{results}. The research team might want to know: given the sampling variability in the new study, how much concern needs to be taken over the fact that the variability introduced by the ocean was ignored? In other words, what statements can be made about the overall risk ratio across all years based on the results from a single year?

In order to answer this question, consider the following framework. Suppose the new study plans to use output from an ensemble of climate models, say $X_{t1}, \dots, X_{tn}$, in order to estimate the risk ratio. Define the new study's estimate of the (log) risk ratio to be $
\widehat{\xi}_t \equiv \widehat{\xi}(X_{t1}, \dots, X_{tn})
$; the ``$t$'' subscript denotes that the estimator is for the risk ratio in a specific year or time period $t$. Whatever the functional form of the estimator, we might assume that its sampling distribution is something like $\widehat{\xi}_t \sim N(\xi_t, \nu^2/n)$; that is, the sampling distribution of the estimator $\widehat{\xi}_t$ is Gaussian and centered on $\xi_t$, the true log risk ratio for year $t$, with the uncertainty introduced by finite sampling represented by the variance $\nu^2/n$. For large enough $n$, the central limit theorem ensures that the Gaussian assumption provides a good approximation.

As the notation implies, a confidence interval for $\xi_t$ using only $\widehat{\xi}_t$ and $\nu^2/n$ is in fact only an interval estimate for the (log) risk ratio in a single year. In fact, the true log risk ratio for year $t$ (${\xi}_t$) itself comes from a population distribution which involves variability, in this case arising from the unaccounted-for effect of the ocean's internal variability. That is, we might have something like $\xi_t \sim N(\mu, \sigma^2)$, where now $\mu$ represents a ``population'' mean (the population being all possible years) and $\sigma^2$ represents the magnitude of oceanic internal variability (for events of a certain type in the specified region). Note that the $\sigma^2$ here is the same quantity as the one introduced in Section \ref{methods}.

Recall that we wish to make a statement about the overall risk ratio across all years based on results from a single year: i.e., use $\widehat{\xi}_t$ to estimate the population distribution $N(\mu, \sigma^2)$. 
Statistical theory allows us to derive the sampling distribution of $\widehat{\xi}_t$ conditional on $\mu$ and the variance components, which is
\begin{equation} \label{newSD}
\widehat{\xi}_t | \mu \sim N(\mu, \nu^2/n + \sigma^2).
\end{equation}
(for details, see Appendix \ref{CIdetails}). In what follows, we assume that the sampling variance $\nu^2$ and additional variance component $\sigma^2$ are known. In fact, these components are of course unknown and must be estimated: $\nu^2$ based on the form of the estimator, and $\sigma^2$ from the statistical model we propose in this paper. This assumption ignores any possible uncertainty in these estimates and may be inappropriate: we emphasize that this is a very rough approximation. Other assumptions might further compromise the following approach, for example assuming that the population distribution for $\xi_t$ is also Gaussian.

In any case, using this sampling distribution, we are equipped to estimate features of the population distribution, based on $\widehat{\xi}_t$. Specifically, we might be interested in estimating a particular percentile of (\ref{newSD}), namely
\[
\phi_p =  \mu + c_p\sqrt{\nu^2/n + \sigma^2},
\]
where the critical value $c_p$ is such that $\mathbbm{P}(Y \leq c_p) = p$, where $Y$ is a standard Gaussian random variable (with mean 0 and variance 1). Note that $\phi_{0.5}$ corresponds to a confidence interval for $\mu$. A $100(1-\alpha)\%$ confidence interval for $\phi_p$ is then
\begin{equation} \label{newCI}
\left( \widehat{\xi}_t + (c_p - z_{\alpha/2}) \sqrt{\nu^2/n + \sigma^2}, \widehat{\xi}_t + (c_p + z_{\alpha/2}) \sqrt{\nu^2/n + \sigma^2} \right),
\end{equation}
where $z_{\alpha/2}$ the critical value for a $100(1-\alpha)\%$ confidence interval. (Details for the derivation of (\ref{newCI}) are also provided in Appendix \ref{CIdetails}.) For example, a 95\% confidence interval for the lower 5th percentile of the population distribution, i.e., $\phi_{0.05} = \mu - 1.645\sqrt{\nu^2/n + \sigma^2}$, is
\begin{equation*} 
\left( \widehat{\xi}_t + (-1.645 - 1.96) \sqrt{\nu^2/n + \sigma^2}, \widehat{\xi}_t + (-1.645 + 1.96) \sqrt{\nu^2/n + \sigma^2} \right)
\end{equation*}
or
\begin{equation} \label{CI005} 
\left( \widehat{\xi}_t -3.605 \sqrt{\nu^2/n + \sigma^2}, \widehat{\xi}_t + 0.315 \sqrt{\nu^2/n + \sigma^2} \right).
\end{equation}

The main idea is that this formula could be used by the hypothetical research team to make a statement about the population distribution of risk values over all possible years, based on a single year. The research team would provide values of $\widehat{\xi}_t$ and $\nu^2/n$, but could then refer to this paper for an estimate of $\sigma^2$ that corresponds to the correct geographic region and event type. If the research team finds that the (log) risk ratio for a particular year is significantly larger than $\log 1 = 0$, then they might wish to estimate whether a lower percentile of the population distribution, e.g. $\phi_{0.05}$, is also larger than $\log 1 = 0$. If so (i.e., if the interval in (\ref{CI005}) is entirely above zero), then the team would have evidence that their results are robust to the fact that they only considered a single year in their study.

\singlespacing
\section*{Acknowledgements}
This work was supported by the Regional and Global Climate Modeling Program of the Office of Biological and Environmental Research in the Department of Energy Office of Science under contract number DE-AC02-05CH11231. This document was prepared as an account of work sponsored by the United States Government. While this document is believed to contain correct information, neither the United States Government nor any agency thereof, nor the Regents of the University of California, nor any of their employees, makes any warranty, express or implied, or assumes any legal responsibility for the accuracy, completeness, or usefulness of any information, apparatus, product, or process disclosed, or represents that its use would not infringe privately owned rights. Reference herein to any specific commercial product, process, or service by its trade name, trademark, manufacturer, or otherwise, does not necessarily constitute or imply its endorsement, recommendation, or favoring by the United States Government or any agency thereof, or the Regents of the University of California. The views and opinions of authors expressed herein do not necessarily state or reflect those of the United States Government or any agency thereof or the Regents of the University of California.

\clearpage
\singlespacing
\bibliographystyle{apalike} \bibliography{risk_ratio_over_time}

\begin{thebibliography}{}

\bibitem[NAS, 2016]{NAS_2016}
 (2016).
\newblock {\em Attribution of extreme weather events in the context of climate
  change}.
\newblock The National Academies Press.

\bibitem[Allen, 2003]{AllenM_2003}
Allen, M. (2003).
\newblock Liability for climate change.
\newblock {\em Nature}, 421:891--892.

\bibitem[Ang\'elil et~al., 2016]{AngelilO_StoneD_etalii_2016}
Ang\'elil, O., Stone, D., Wehner, M., Paciorek, C.~J., Krishnan, H., and
  Collins, W. (2016).
\newblock An independent assessment of anthropogenic attribution statements for
  recent extreme temperature and rainfall events.
\newblock {\em J. Climate}, page In press.

\bibitem[Ang\'elil et~al., 2014a]{AngelilO_StoneDA_PallP_2014}
Ang\'elil, O., Stone, D.~A., and Pall, P. (2014a).
\newblock Attributing the probability of {S}outh {A}frican weather extremes to
  anthropogenic greenhouse gas emissions: spatial characteristics.
\newblock {\em Geophy. Res. Lett.}, 41:3238--3243.

\bibitem[Ang\'elil et~al., 2014b]{AngelilO_StoneDA_etalii_2014}
Ang\'elil, O., Stone, D.~A., Tadross, M., Tummon, F., Wehner, M., and Knutti,
  R. (2014b).
\newblock Attribution of extreme weather to anthropogenic greenhouse gas
  emissions: sensitivity to spatial and temporal scales.
\newblock {\em Geophys. Res. Lett.}, 41:2150--2155.

\bibitem[DerSimonian and Laird, 1986]{Dersimonian1986}
DerSimonian, R. and Laird, N. (1986).
\newblock Meta-analysis in clinical trials.
\newblock {\em Controlled Clinical Trials}, 7(3):177 -- 188.

\bibitem[Gelman et~al., 2013]{GelmanBDA}
Gelman, A., Carlin, J.~B., Stern, H.~S., Dunson, D.~B., Vehtari, A., and Rubin,
  D.~B. (2013).
\newblock {\em {Bayesian Data Analysis}}.
\newblock Chapman and Hall/CRC, 3rd edition.

\bibitem[Hansen et~al., 2014]{HansenG_AuffhammerM_SolowAR_2014}
Hansen, G., Auffhammer, M., and Solow, A.~R. (2014).
\newblock On the attribution of a single event to climate change.
\newblock {\em J. Climate}, 27:8297--8301.

\bibitem[Hurrell et~al., 2008]{HurrellJW_HackJJ_etalii_2008}
Hurrell, J.~W., Hack, J.~J., Shea, D., Caron, J.~M., and Rosinski, J. (2008).
\newblock A new sea surface temperature and sea ice boundary dataset for the
  {C}ommunity {A}tmosphere {M}odel.
\newblock {\em J. Climate}, 21:5145--5153.

\bibitem[Jeon et~al., 2016]{Jeon2016}
Jeon, S., Paciorek, C.~J., and Wehner, M.~F. (2016).
\newblock Quantile-based bias correction and uncertainty quantification of
  extreme event attribution statements.
\newblock {\em Weather and Climate Extremes}, pages~--.

\bibitem[McCulloch and Neuhaus, 2005]{glmm}
McCulloch, C.~E. and Neuhaus, J.~M. (2005).
\newblock {\em Generalized Linear Mixed Models}.
\newblock John Wiley \& Sons, Ltd.

\bibitem[Neale et~al., 2012]{NealeRB_ChenC-C_etalii_2012}
Neale, R.~B., Chen, C.-C., Gettelman, A., Lauritzen, P.~H., Park, S.,
  Williamson, D.~L., Conley, A.~J., Garcia, R., Kinnison, D.~Lamarque, J.-F.,
  Marsh, D., Mills, M., Smith, A.~K., Tilmes, S.~Vitt, F., Morrison, H.,
  Cameron-Smith, P., Collins, W.~D., Iacono, M.~J., Easter, R.~C., Ghan, S.~J.,
  Liu, X., Rasch, P.~J., and Taylor, M.~A. (2012).
\newblock Description of the {NCAR} community atmosphere model ({CAM~5.0}).
\newblock Technical report, NCAR Technical Note NCAR/TN-486+STR.

\bibitem[Otto et~al., 2015]{OttoFEL_BoydE_etalii_2015}
Otto, F. E.~L., Boyd, E., Jones, R.~G., Cornforth, R.~J., James, R., Parker,
  H.~R., and Allen, M.~R. (2015).
\newblock Attribution of extreme weather events in {A}frica: a preliminary
  exploration of the science and policy implications.
\newblock {\em Clim. Change}, 132:531--543.

\bibitem[Pall et~al., 2011]{PallP_AinaT_etalii_2011}
Pall, P., Aina, T., Stone, D.~A., Stott, P.~A., Nozawa, T., Hilberts, A. G.~J.,
  Lohmann, D., and Allen, M.~R. (2011).
\newblock Anthropogenic greenhouse gas contribution to flood risk in {E}ngland
  and {W}ales in {A}utumn 2000.
\newblock {\em Nature}, 470:382--385.

\bibitem[Stone and Allen, 2005]{StoneDA_AllenMR_2005a}
Stone, D.~A. and Allen, M.~R. (2005).
\newblock The end-to-end attribution problem: from emissions to impacts.
\newblock {\em Clim. Change}, 71:303--318.

\bibitem[Stone and Pall, 2016]{StoneDA_PallP_2016}
Stone, D.~A. and Pall, P. (2016).
\newblock A benchmark estimate of the effect of anthropogenic emissions on the
  ocean surface.
\newblock page In prep.

\bibitem[Trenberth et~al., 2007]{IPCC2007}
Trenberth, K., Jones, P., Ambenje, P., Bojariu, R., Easterling, D., Tank,
  A.~K., Parker, D., Rahimzadeh, F., Renwick, J., Rusticucci, M., Soden, B.,
  and Zhai, P. (2007).
\newblock {\em {Observations: Surface and Atmospheric Climate Change. In:
  Climate Change 2007: The Physical Science Basis. Contribution of Working
  Group I to the Fourth Assessment Report of the Intergovernmental Panel on
  Climate Change}}.
\newblock Cambridge University Press.

\bibitem[Wolski et~al., 2014]{WolskiP_StoneD_etalii_2014}
Wolski, P., Stone, D., Tadross, M., Wehner, M., and Hewitson, B. (2014).
\newblock Attribution of floods in the {O}kavango {B}asin, {S}outhern {A}frica.
\newblock {\em J. Hydrol.}, 511:350--358.

\end{thebibliography}

\clearpage
\begin{appendix}
\numberwithin{equation}{section}

\section{Details for the derivation of the confidence interval of $\phi_p$} \label{CIdetails}

Recall the setting introduced in Section \ref{discussion}, the goal being to make a confidence statement regarding the population distribution of risk from all possible years, using only a risk estimate from a single year.

The sampling distribution of $\widehat{\xi}_t$ conditional on $\mu$ is derived as follows. Again, recall that the sampling distribution of $\widehat{\xi}_t$ conditional on $\xi_t$ is $N(\xi_t, \nu^2/n)$; also $\xi_t \sim N(\mu, \sigma^2)$. The sampling distribution of interest is calculated by averaging over $\xi_t$:
\[
p(\widehat{\xi}_t | \mu ) = \int_{\xi_t} p(\widehat{\xi}_t | \xi_t) p(\xi_t | \mu) d\xi_t,
\]
where the implicit conditioning on $\nu^2$ and $\sigma^2$ is suppressed in the notation. Given that $p(\widehat{\xi}_t | \xi_t) = N(\xi_t, \nu^2/n)$ and $p(\xi_t | \mu) = N(\mu, \sigma^2)$, the closed-form solution is well-known (this setup is identical to the derivation for the marginal distribution of the data in a Normal-Normal Bayesian posterior calculation). The result is that
\begin{equation} \label{newSampDist}
p(\widehat{\xi}_t | \mu ) = N(\mu, \nu^2/n + \sigma^2).
\end{equation}

Next, to derive a confidence interval for $\phi_p$, first note that using (\ref{newSampDist}) we can obtain a $100(1-\alpha)\%$ confidence interval for $\mu$ as
\[
\left( \widehat{\xi}_t - z_{\alpha/2} \sqrt{\nu^2/n + \sigma^2},  \widehat{\xi}_t + z_{\alpha/2} \sqrt{\nu^2/n + \sigma^2}\right).
\]
Because $\phi_p =  f(\mu) = \mu + c_p\sqrt{\nu^2/n + \sigma^2}$ is a linear function of $\mu$, statistical theory says that $f(\widehat{\xi}_t) \sim N\big(f(\mu), \text{Var}[f(\widehat{\xi}_t)]\big)$, so that a $100(1-\alpha)\%$ confidence interval for $\phi_p = f(\mu)$ is
\[
\left( f(\widehat{\xi}_t) - z_{\alpha/2}\sqrt{\text{Var}[f(\widehat{\xi}_t)]},  f(\widehat{\xi}_t) + z_{\alpha/2}\sqrt{\text{Var}[f(\widehat{\xi}_t)]} \right).
\]
Since $\text{Var}[f(\widehat{\xi}_t)] = \text{Var}\hskip0.5ex \widehat{\xi}_t = \nu^2/n + \sigma^2$, the confidence interval for $\phi_p$ is
\[
\left( \left[\widehat{\xi}_t - c_p\sqrt{\nu^2/n + \sigma^2}\right] - z_{\alpha/2}\sqrt{\nu^2/n + \sigma^2},  \left[\widehat{\xi}_t - c_p\sqrt{\nu^2/n + \sigma^2}\right] + z_{\alpha/2}\sqrt{\nu^2/n + \sigma^2} \right)
\]
or
\[
\left( \widehat{\xi}_t + (c_p - z_{\alpha/2}) \sqrt{\nu^2/n + \sigma^2}, \widehat{\xi}_t + (c_p + z_{\alpha/2}) \sqrt{\nu^2/n + \sigma^2} \right),
\]
which is what is given in (\ref{newCI}).

\clearpage

\section{Supplemental Figures} \label{suppFig}

\begin{figure}[!h]
\begin{center}
\includegraphics[width=\textwidth]{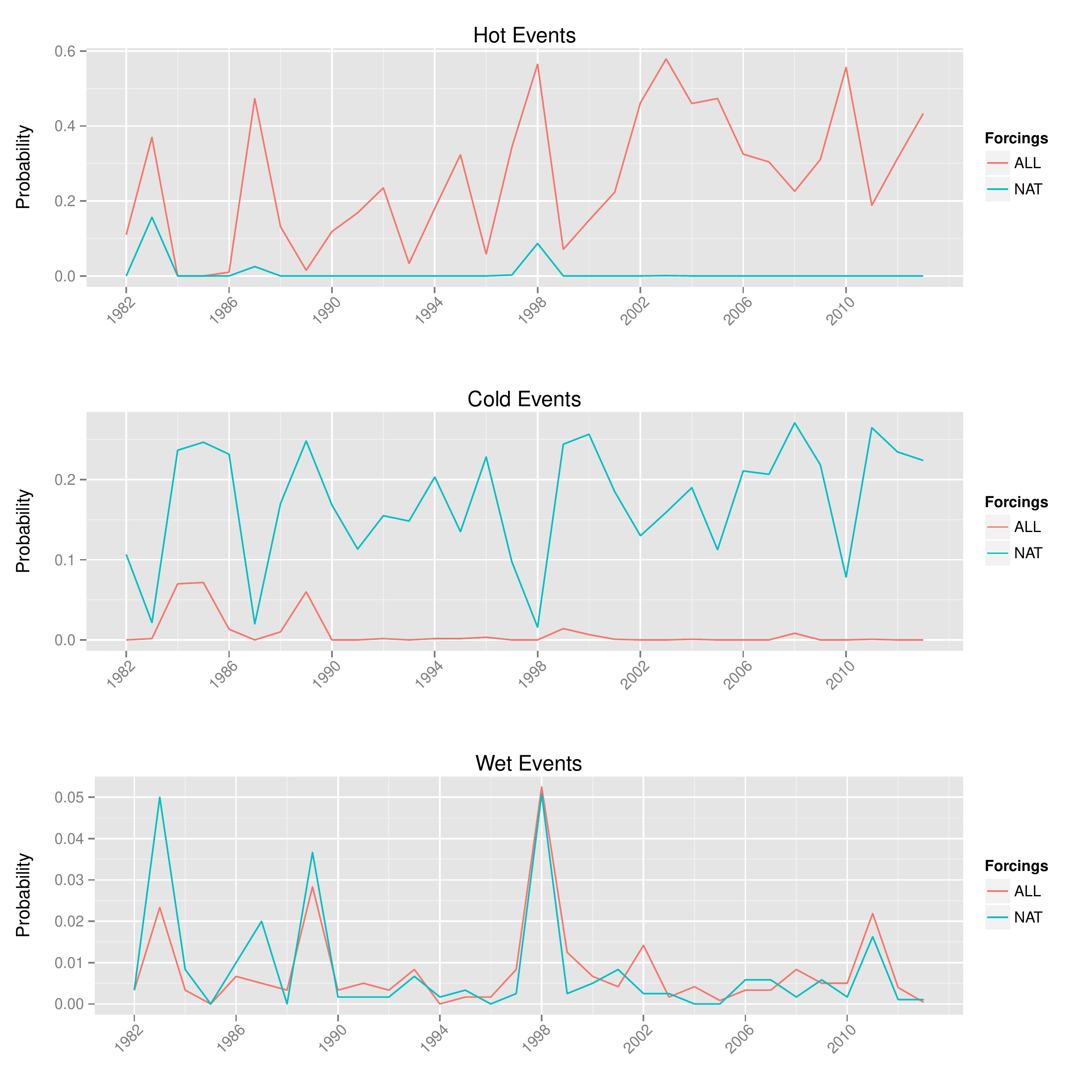}
\caption{Empirical event probabilities averaged across each year for the southern Andean Community.}
\label{sacProbs}
\end{center}
\end{figure}

\begin{figure}[!h]
\begin{center}
\includegraphics[width=\textwidth]{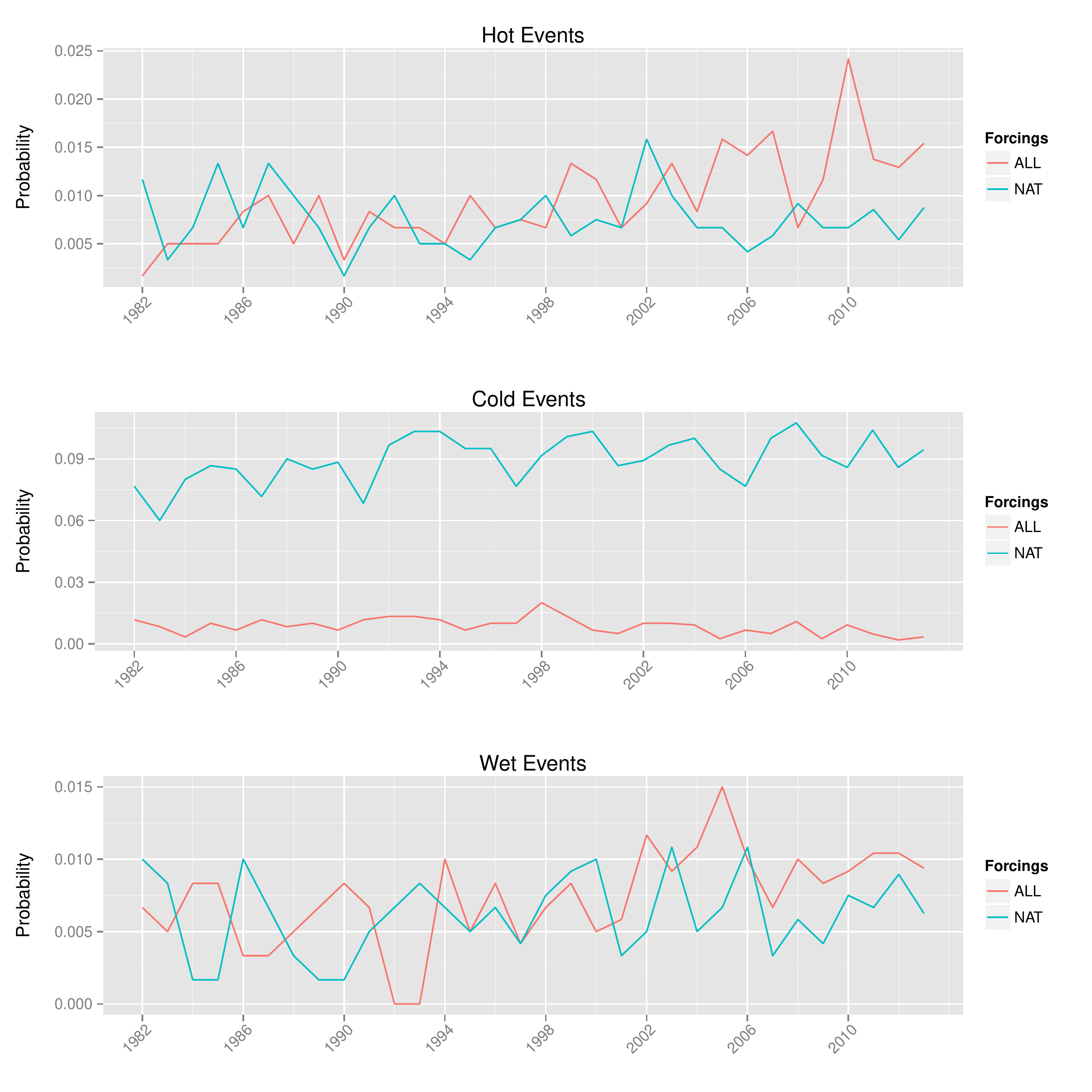}
\caption{Empirical event probabilities averaged across each year for Krasnoyarsk, Russia.}
\label{KrasProbs}
\end{center}
\end{figure}

\begin{figure}[!h]
\begin{center}
\includegraphics[width=\textwidth]{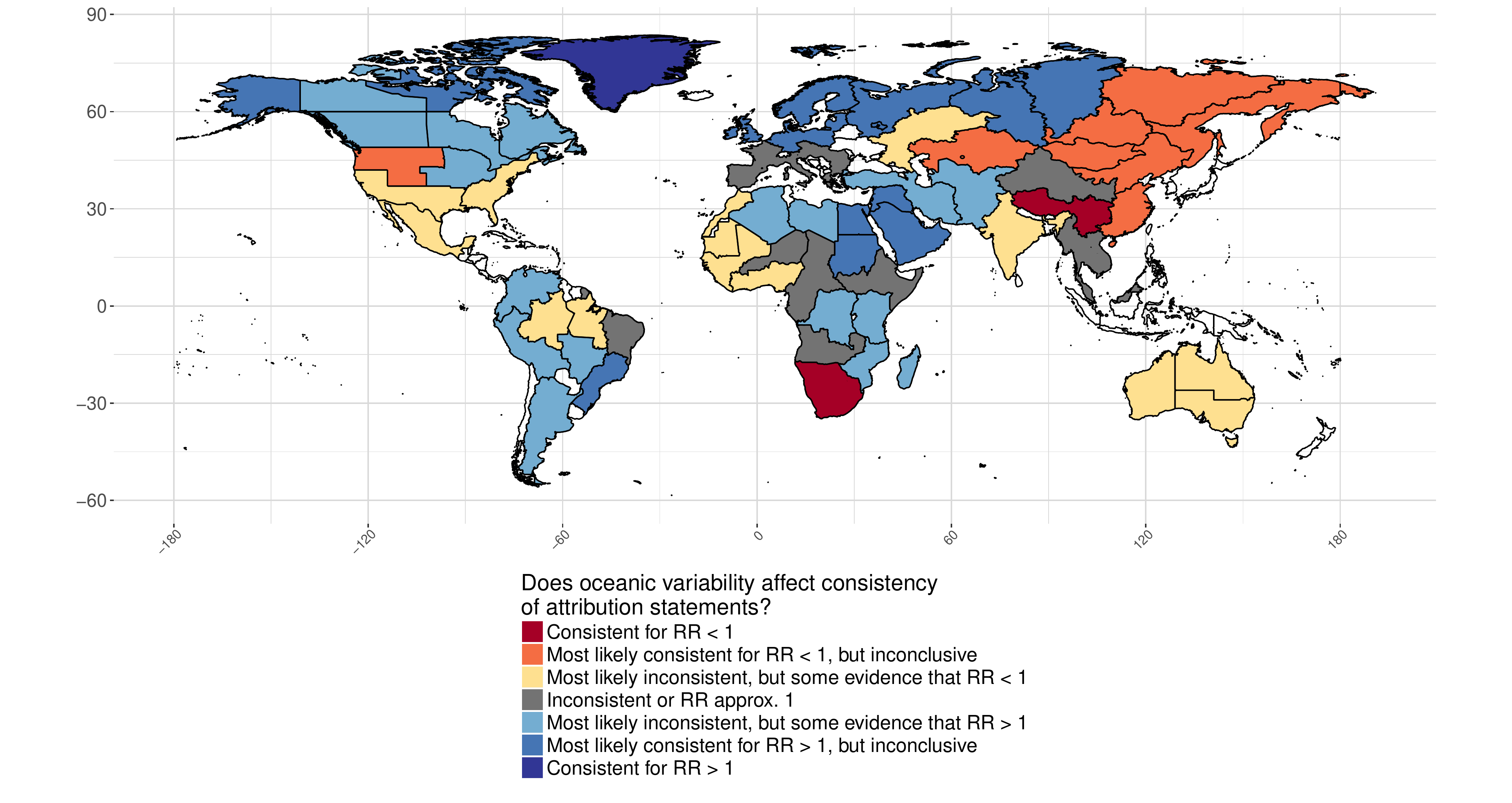}
\caption{Map representation of the color codings for wet events in Figures \ref{wet_EP} and \ref{wetIV}. While there is no relationship between latitude and the colors, there is noticeable clustering.}
\label{wetMap}
\end{center}
\end{figure}

\clearpage

\section{Prior specification} \label{priorSpec}

In a Bayesian framework, all unknown parameters must be assigned a prior distribution, which summarizes all knowledge about the parameters \textit{a priori} to observing data. While in general any known prior information relating to the parameters can be incorporated by way of the prior distribution, we instead use non-informative and proper prior distributions to avoid introducing any biases, meaning that the priors (1) specify essentially no information about the parameters, but (2) are proper statistical distributions.

Recall that the full parameter vector for our model is $\bftheta = (\boldsymbol{\alpha}, \boldsymbol{\delta}, \boldsymbol{\gamma}, \bfbeta_A, \bfbeta_N, \tau^2, \sigma^2, \omega^2)$; we now define the prior distribution $p(\bftheta)$. Note that several components of the prior distribution are specified in the third level of the model, and we can thus decompose the prior as 
\begin{equation} \label{decomp_prior}
p(\bftheta) = p( \boldsymbol{\alpha} | \tau^2) \cdot p( \boldsymbol{\delta} | \sigma^2) \cdot p( \boldsymbol{\gamma} | \omega^2) \cdot  p(\tau^2 )  \cdot p( \sigma^2)  \cdot p(\omega^2 )  \cdot p( \bfbeta_A)  \cdot p( \bfbeta_N).
\end{equation}
Throughout, the vertical bar ``$|$'' means ``conditional on.'' Each of these components are as follows:
\[
\begin{array}{c}
p( \boldsymbol{\alpha} | \tau^2) = N_T({\bf 0}, \tau^2{\bf I}_{T}), \hskip2ex p( \boldsymbol{\delta} | \sigma^2) = N_T({\bf 0}, \sigma^2{\bf I}_{T}), \hskip2ex
p( \boldsymbol{\gamma} | \omega^2) = N_{12}({\bf 0}, \omega^2{\bf I}_{12}) \cdot \mathbbm{1}_{\{ \sum_{j} \gamma_j = 0 \} }, \\ [1ex]
p(\tau^2 ) =U(0, 1000), \hskip2ex p( \sigma^2)= U(0, 1000), \hskip2ex p(\omega^2 ) = C(10),  \\[1ex]
p( \bfbeta_A) = N_{2}({\bf 0}, 10^2{\bf I}_{2}) , \hskip2ex p( \bfbeta_N) = N_{2}({\bf 0}, 10^2{\bf I}_{2})
\end{array}
\]
Here, $N_q({\bf a}, {\bf B})$ is a $q$-variate Gaussian distribution with mean ${\bf a}$ and covariance ${\bf B}$; $U(c, d)$ is the uniform distribution over the interval $(c, d)$; $C(s)$ is the half-Cauchy distribution, i.e., the positive portion of a Cauchy distribution centered on 0 and with scale parameter $s$. Note: the prior on $\boldsymbol{\gamma}$ imposes the restriction that the empirical mean is zero.

One additional constraint is imposed on the prior, in order to account for a problem with the mixing of the Markov chain Monte Carlo (MCMC) algorithm for several of the event types and regions. The issue has to do with the fact that the monthly effects are defined on the $\logit$ scale, and therefore the model can have trouble distinguishing between $\logit^{-1}(-10)$ and $\logit^{-1}(-100)$ since these are both essentially zero. This leads to a pseudo ``non-identifiability'' when many of the monthly probabilities are zero; in other words, when the seasonality is such that essentially all of the extreme events occur in a single calendar month (e.g., for hot events in the northern extra-tropics, almost all hot events occur in July). In these cases, some of the monthly effects ($\gamma_j$) trade off with the scenario-specific intercepts ($\beta_{A0}$ and $\beta_{N0}$): these parameters mix very poorly, while trace plots for the probabilities themselves indicate convergence for the MCMC.

To account for this, for events where this poor mixing arises we add an additional constraint on the prior: we require the logit monthly probabilities to lie between $\pm L$, where the constant $L$ is chosen separately for each region and event type to ensure that the $\gamma_j$ and $\beta_{k0}$ mix properly (often, $L = 10$ to $20$). Formally, the prior we use is
\begin{equation} \label{final_prior}
\widetilde{p}(\bftheta) = p(\bftheta) \cdot \mathbbm{1}_{\big\{ \min \{\logit p_{ktj}\} > -L \big\}} \cdot \mathbbm{1}_{\big\{ \max \{\logit p_{ktj}\} <L \big\}},
\end{equation}
where $p(\bftheta)$ is from (\ref{decomp_prior}).

\section{Markov chain Monte Carlo} \label{MCMC}

The hierarchical statistical model introduced in Section 3 of the main paper is estimated using a Bayesian paradigm. As is often the case, the posterior distribution given in Equation (5) is not available in closed form, and we therefore resort to Markov chain Monte Carlo (MCMC) methods (for a summary see, e.g., Gilks, Richardson, and Spiegelhalter, 1996) to obtain joint samples from the posterior distribution (i.e., the distribution of all parameters conditional on observed data). However, the nature of our model requires several computational tools above and beyond standard MCMC methods (e.g., Gibbs sampling and Metropolis-Hastings), which are now described.

First, recall that we defined random variables ${\bf Z} = \{ Z_{ktj}:  k \in \{A,N\}; t=1, \dots, T; j = 1, \dots, 12 \}$; now define ${\bf z}$ to represent the observed values of ${\bf Z}$. For clarity, we rewrite the likelihood for each $Z_{ktj}$ in terms of $\bftheta$ as
\begin{equation} \label{indiv_likelihood}
p(Z_{ktj} | \bftheta) = \text{binomial}\big(n_t, \logit^{-1}[{\bf x}_{kt}^\top\bfbeta_k + \alpha_t + \delta_t \mathbbm{1}_{\{k=A\}} + \gamma_j ] \big),
\end{equation}
and the likelihood for all $Z_{ktj}$ is
\begin{equation} \label{likelihood}
p( {\bf Z} | \bftheta) =  \prod_{k \in \{A, N\}} \prod_{t=1}^T \prod_{j=1}^{12} p(Z_{ktj} | \bftheta).
\end{equation}
Then, the posterior distribution of interest is
\begin{equation} \label{post}
p(\bftheta | {\bf Z =  z}) = \frac{p({\bf z} | \bftheta) \widetilde{p}(\bftheta)}{ \int_\bftheta p({\bf z} | \bftheta) \widetilde{p}(\bftheta) d\bftheta} \propto p({\bf z} | \bftheta) \widetilde{p}(\bftheta),
\end{equation}
where $p({\bf z} | \bftheta)$ is from (\ref{likelihood}) and $\widetilde{p}(\bftheta)$ is the prior distribution from (\ref{final_prior}). Several strategies for sampling from (\ref{post}) in a MCMC framework are as follows.

\subsection{Traditional Gibbs sampler with Metropolis-Hastings}

A traditional MCMC approach to sampling from (\ref{post}) is as follows:

\noindent\hrulefill

\noindent {\rm {\bf Traditional Sampler:} }

\begin{itemize}
\item[\textit{Step 1}:] Draw $(\boldsymbol{\alpha}, \boldsymbol{\delta})$ from $p(\boldsymbol{\alpha}, \boldsymbol{\delta} |  \boldsymbol{\gamma}, \bfbeta_A, \bfbeta_N, \tau^2, \sigma^2, \omega^2, {\bf Z} ) = p(\boldsymbol{\alpha}, \boldsymbol{\delta} |  \boldsymbol{\gamma}, \bfbeta_A, \bfbeta_N, {\bf Z} )$

\item[\textit{Step 2}:] Draw $\boldsymbol{\gamma}$ from $p( \boldsymbol{\gamma} | \boldsymbol{\alpha}, \boldsymbol{\delta}, \bfbeta_A, \bfbeta_N, \tau^2, \sigma^2, \omega^2, {\bf Z} ) = p( \boldsymbol{\gamma} | \boldsymbol{\alpha}, \boldsymbol{\delta}, \bfbeta_A, \bfbeta_N, {\bf Z} )$

\item[\textit{Step 3 (A)}:] Draw $( \bfbeta_A, \bfbeta_N)$ from $p( \bfbeta_A, \bfbeta_N | \boldsymbol{\alpha}, \boldsymbol{\delta},  \boldsymbol{\gamma}, \tau^2, \sigma^2, \omega^2, {\bf Z} ) = p( \bfbeta_A, \bfbeta_N | \boldsymbol{\alpha}, \boldsymbol{\delta}, \boldsymbol{\gamma}, {\bf Z} )$

\item[\textit{Step 4 (S)}:] (a) Draw $\tau^2$ from $p( \tau^2 | \bfbeta_A, \bfbeta_N, \boldsymbol{\alpha}, \boldsymbol{\delta}, \sigma^2,  \omega^2, {\bf Z} ) = p( \tau^2 | \boldsymbol{\alpha} )$

(b) Draw $\sigma^2$ from $p( \sigma^2 | \bfbeta_A, \bfbeta_N, \boldsymbol{\alpha}, \boldsymbol{\delta}, \tau^2, \omega^2, {\bf Z} ) = p( \sigma^2 | \boldsymbol{\delta} )$

\item[\textit{Step 5}:] Draw $\omega^2$ from $p( \omega^2 |\boldsymbol{\alpha}, \boldsymbol{\delta}, \boldsymbol{\gamma},  \bfbeta_A, \bfbeta_N, \tau^2, \sigma^2, {\bf Z} ) = p( \omega^2 | \boldsymbol{\gamma} )$

\end{itemize}

\vskip-2ex

\noindent\hrulefill

\noindent Regardless of prior choice, steps 1-3 will require Metropolis-Hastings; while choosing conjugate priors for $\tau^2$, $\sigma^2$, and $\omega^2$ could allow for closed-form Gibbs updates in steps 4 (a)-(b) and 5, the non-informative uniform priors used (see Supplemental Section \ref{priorSpec}) also necessitate Metropolis-Hastings steps. The \textit{(A)} and \textit{(S)} labels will be explained in the next section.

\subsection{Ancillarity-Sufficiency Interweaving Strategy (ASIS) MCMC}

The problem with using the Traditional Sampler defined above lies in the fact that the regression coefficients $\bfbeta_k$ and yearly effects $(\boldsymbol{\alpha}, \boldsymbol{\delta})$ are highly correlated, resulting in extremely poor mixing of the Markov chain. In order to fix this problem, we use the Ancillarity-Sufficiency Interweaving Strategy (ASIS) method of Yu and Meng (2011). MCMC for multilevel models can be set up using either a centered parameterization (CP) or a non-centered parameterization (NCP); their approach proposes a strategy that alternates (``interweaves'') sampling from both parameterizations. The ASIS terminology replaces the CP/NCP terminology by framing the problem in terms of the mathematically equivalent data augmentation schemes of \textit{ancillary augmentation} (AA) and \textit{sufficient augmentation} (SA), arguing that this terminology better captures the ``essence'' of the method. Regardless, in the AA scheme (i.e., NCP), the missing data (i.e., latent variables) are an ancillary statistic for the parameter of interest, and in the SA scheme (i.e., CP), the missing data are a sufficient statistic for the parameter of interest.

Note now that the parameterization of our model for $\boldsymbol{\alpha}$ and $\boldsymbol{\delta}$ are AA for $( \bfbeta_A, \bfbeta_N)$ but SA for $( \tau^2, \sigma^2)$ (hence, the \textit{A} and \textit{S} notation in the Traditional sampler). To implement the ASIS sampler, we need to find an AA for $( \tau^2, \sigma^2)$ and an SA for $( \bfbeta_A, \bfbeta_N)$.

\vskip3ex
\noindent \underline{SA for $( \bfbeta_A, \bfbeta_N)$:} 

\noindent Define new latent variables $\eta_t = {\bf x}_{Nt}^\top\bfbeta_N + \alpha_t$ and $\nu_t = {\bf x}_{At}^\top\bfbeta_A + \alpha_t + \delta_t$, so that the parameterization in (\ref{indiv_likelihood}) can be rewritten as
\begin{equation} \label{model2a}
p(Z_{ktj} | \bftheta) = \text{binomial}\big(n_t, \logit^{-1}[\eta_t \mathbbm{1}_{\{k=N\}} + \nu_t \mathbbm{1}_{\{k=A\}} + \gamma_j ] \big),
\end{equation}
where now
\begin{equation} \label{model2b}
\eta_t \stackrel{\text{ind}}{\sim} N({\bf x}_{Nt}^\top\bfbeta_N, \tau^2) \hskip3ex \text{and} \hskip3ex \nu_{t} \stackrel{\text{ind}}{\sim} N(  {\bf x}_{At}^\top\bfbeta_A, \tau^2 + \sigma^2).
\end{equation}
Note that $p({\bf Z} | \boldsymbol{\eta}, \boldsymbol{\nu}, \boldsymbol{\gamma}, \bfbeta_A, \bfbeta_N, \tau^2, \sigma^2) = p({\bf Z} | \boldsymbol{\eta}, \boldsymbol{\nu}, \boldsymbol{\gamma})$; therefore $(\boldsymbol{\eta}, \boldsymbol{\nu})$ are SA for $( \bfbeta_A, \bfbeta_N)$ (and, for that matter, also for $( \tau^2, \sigma^2)$).

The traditional sampler can be improved by adding a step that samples $( \bfbeta_A, \bfbeta_N)$ under the SA. Actually, this step will be split into two steps so that Gibbs steps can be used:

\begin{itemize}
\item[\textit{Step 3 (S)}:] (a) Draw $\bfbeta_A$ from $p( \bfbeta_A | \bfbeta_N,  \boldsymbol{\eta}, \boldsymbol{\nu}, \boldsymbol{\gamma}, \tau^2, \sigma^2, \omega^2, {\bf Z} ) = p( \bfbeta_A | \boldsymbol{\nu}, \tau^2, \sigma^2 )$

(b) Draw $\bfbeta_N$ from $p( \bfbeta_N | \bfbeta_A,  \boldsymbol{\eta}, \boldsymbol{\nu}, \boldsymbol{\gamma}, \tau^2, \sigma^2, \omega^2, {\bf Z} ) = p( \bfbeta_N | \boldsymbol{\eta}, \tau^2 )$

\end{itemize}

\noindent Using conjugate (Gaussian) priors for $\bfbeta_N$ and $\bfbeta_A$, these can be sampled in a Gibbs step.

\vskip3ex
\noindent \underline{AA for $( \tau^2, \sigma^2)$:} 

\noindent Define new latent variables $\kappa_t = \alpha_t/\tau$ and $\xi_t = \delta_t/\sigma$; the parameterization in (\ref{indiv_likelihood}) can be rewritten as
\begin{equation} \label{model3a}
p(Z_{ktj} | \bftheta) = \text{binomial}\big(n_t, \logit^{-1}[{\bf x}_{kt}^\top\bfbeta_k + \tau \kappa_t + \sigma \xi_t \mathbbm{1}_{\{k=A\}} + \gamma_j ] \big)
\end{equation}
and
\begin{equation} \label{model3b}
\kappa_t \stackrel{\text{iid}}{\sim} N(0, 1) \hskip3ex \text{and} \hskip3ex \xi_{t} \stackrel{\text{iid}}{\sim} N(  0, 1).
\end{equation}
Because of (\ref{model3b}), we can see that $(\boldsymbol{\kappa}, \boldsymbol{\xi})$ are  AA for $( \tau^2, \sigma^2)$ (and $( \bfbeta_A, \bfbeta_N)$). Hence, we can add the following step:

\begin{itemize}
\item[\textit{Step 4 (A)}:] Draw $( \tau^2, \sigma^2 )$ from $p(  \tau^2, \sigma^2 |  \boldsymbol{\gamma}, \bfbeta_A , \bfbeta_N,  \boldsymbol{\kappa}, \boldsymbol{\xi}, {\bf Z} )$

\end{itemize}

\noindent This step again requires Metropolis-Hastings.

Combining all of the above, the following is a componentwise interweaving sampler using ASIS. The intermediate latent variables are calculated between the $A$ and $S$ step, and the original variables are then updated after the $S$ step. For each of these calculations, the most recently sampled parameter values are used: the ``$*$'' sub- and superscripts are used to explicitly show this. Specific details on how each step is carried out are also provided; for example, many of the yearly and monthly effects are sampled sequentially.

\noindent\hrulefill

\noindent {\rm {\bf Componentwise Interweaving Sampler:} }

\begin{itemize}

\item[\textit{Step 1}:] For $t=1, \dots, T$, draw $(\alpha_t^*, \delta_t^*)$ from $p(\alpha_t, \delta_t | \boldsymbol{\gamma}, \bfbeta_A, \bfbeta_N, {\bf Z} )$

{\small 
For each $t$, this involves a random-walk Metropolis-Hastings (RWMH) step. The bivariate proposal distribution is centered on current values, and the proposal correlation is fixed to a large negative number ($\approx -0.95$) to improve mixing.
}

\item[\textit{Step 2}:] For $j = 1, \dots, 12$, draw $\gamma^*_j$ from $p( \gamma_j | \boldsymbol{\alpha}^*, \boldsymbol{\delta}^*, \bfbeta_A, \bfbeta_N, {\bf Z} )$

{\small 
For each $j$, this again involves a RWMH step. Recall the prior constraint that $\sum_{j=1}^{12} \gamma_j = 0$; this constraint is imposted as follows:
first, a new component $\gamma_j^{\text{prop}}$ is proposed, and define $\widetilde{\boldsymbol{\gamma}} = (\gamma_1^{\text{curr}}, \dots, \gamma_j^{\text{prop}}, \dots, \gamma_{12}^{\text{curr}})$. Then, the other components of $\boldsymbol{\gamma}$ are also adjusted to arrive at a fully adjusted proposal:
\[
\boldsymbol{\gamma}^{\text{prop}} = \widetilde{\boldsymbol{\gamma}} - \frac{1}{12} \sum_{k=1}^{12} \widetilde{{\gamma_k}}.
\]
As long as the initialized value of $\boldsymbol{\gamma}$ is mean-zero, this adjustment will preserve the prior constraint. For each $j$, the entire vector $\boldsymbol{\gamma}^{\text{prop}}$ is either accepted or rejected.
}

\item[\textit{Step 3 (A)}:] Draw $( \bfbeta_A^*, \bfbeta_N^*)$ from $p( \bfbeta_A, \bfbeta_N | \boldsymbol{\alpha}^*, \boldsymbol{\delta}^*, \boldsymbol{\gamma}^*, {\bf Z} )$

{\small 
This step is carried out using RWMH, with a separate bivariate proposal for each $\bfbeta_A$ and $\bfbeta_N$. The within-scenario coefficients are blocked together in order to avoid poor mixing. For wet events, the proposal correlation is 0, as there were no problems with mixing. However, for hot and cold events, there was poor mixing for the intercept and temperature coefficient for the scenario which had essentially all zero counts (NAT for hot events and ALL for cold events). In these cases, the proposal correlation for these scenarios was fixed to a large negative value ($\approx -0.95$).
}

\item[\textbf{Calculate:}]  $\eta_t = {\bf x}_{Nt}^\top\bfbeta_N^* + \alpha_t^*$ and $\nu_t = {\bf x}_{At}^\top\bfbeta_A^* + \alpha_t + \delta_t^*$ for $t=1,\dots, T$.

\item[\textit{Step 3 (S)}:] (a) Draw $\bfbeta_A^{**}$ from $p( \bfbeta_A | \boldsymbol{\nu}, \tau^2, \sigma^2 )$

(b) Draw $\bfbeta_N^{**}$ from $p( \bfbeta_N | \boldsymbol{\eta}, \tau^2 )$

{\small 
Using conjugate Gaussian priors for $\bfbeta_A$ and $\bfbeta_N$, this step can be accomplished with a closed-form Gibbs update.
}

\item[\textbf{Update:}]  $\alpha_t^{**} = \eta_t - {\bf x}_{Nt}^\top\bfbeta_N^{**}$ and $\delta_t^{**} = \nu_t - {\bf x}_{At}^\top\bfbeta_A^{**} - \alpha_t^{**}$ for $t=1,\dots, T$.

\item[\textbf{Calculate:}]  $\kappa_t = \alpha_t^{**}/\tau$ and $\xi_t = \delta_t^{**}/\sigma$ for $t=1,\dots, T$.

\item[\textit{Step 4 (A)}:] Draw $( \tau^2_*, \sigma^2_* )$ from $p(  \tau^2, \sigma^2 |  \boldsymbol{\gamma}^*, \bfbeta_A^{**} , \bfbeta_N^{**},  \boldsymbol{\kappa}, \boldsymbol{\xi}, {\bf Z} )$ (RWMH, separately for $\tau^2$ and $\sigma^2$)

\item[\textbf{Update:}]  $\alpha_t^{***} = \tau_{*}\kappa_t$ and $\delta_t^{***} = \sigma_{*}\xi_t$ for $t=1,\dots, T$.

\item[\textit{Step 4 (S)}:] (a) Draw $\tau^2_{**}$ from $p( \tau^2 | \boldsymbol{\alpha}^{***} )$ (RWMH)

(b) Draw $\sigma^2_{**}$ from $p( \sigma^2 | \boldsymbol{\delta}^{***} )$ (RWMH)

\item[\textit{Step 5}:] Draw $\omega^2_*$ from $p( \omega^2 | \boldsymbol{\gamma}^* )$ (RWMH)

\end{itemize}

\vskip-2ex

\noindent\hrulefill

\noindent At the end of each iteration, save the most recently sampled/calculated values of the parameters: $(\boldsymbol{\alpha}^{***}, \boldsymbol{\delta}^{***}, \boldsymbol{\gamma}^{*}, \bfbeta_A^{**}, \bfbeta_N^{**}, \tau^2_{**}, \sigma^2_{**}, \omega^2_*)$. Note: the proposal variance for each of the RWMH steps was tuned such that the acceptance probability falls between $0.3$ and $0.4$.

\section{Software and data} \label{software}

\begin{table}[!t]
\begin{center}
\begin{tabular}{|l|c|} 
\hline
\textbf{Name} & \textbf{Version} \\ \hline \hline
{\tt R} 					& 3.2.3 \\ \hline \hline
{\tt colorspace} 		& 1.2-6 \\ \hline
{\tt RColorBrewer} 	& 1.1-2 \\ \hline
{\tt gridExtra} 			& 2.0.0 \\ \hline
{\tt ggplot2} 			& 1.0.1 \\ \hline
{\tt MASS} 			& 7.3-45 \\ \hline
\end{tabular}
\end{center}
\caption{Version of {\tt R} and other packages used to generate the results and plots in the main text.}
\label{versions}
\end{table}%

The componentwise interweaving MCMC sampler was coded up in {\tt R}, along with all other functions required to calculate the adjusted risk ratio and exceedance probabilities and make the plots in the paper. The source code, data files, and a script with replication code are publicly available in the GitHub repository \url{http:/bitbucket.org/markdrisser/timerr_package}.

As a note, the version of {\tt R} and various packages used to generate the results in the paper are listed in Supplemental Table \ref{versions}.

\subsection{Data files}

The following data files are included:

\begin{enumerate}
\item Binomial count data

A total of 58 files are provided, one for each region. Each text file (e.g., {\tt USA-C.txt}) contains the binomial count variables for each month ({\tt january}, {\tt february}, \dots, {\tt december}), as well as additional variables indicating {\tt event\char`_type} ({\tt hot}, {\tt cold}, or {\tt wet}), {\tt scenario} ({\tt ALL} or {\tt NAT}), {\tt year}, and {\tt n\char`_sims} (the ensemble size).

\item {\tt gmt.txt}

This file contains four variables {\tt gmtA\char`_raw}, {\tt gmtA}, {\tt gmtN\char`_raw}, and {\tt gmtN}, each of which contains a temperature value from 1982-2013. The {\tt raw} vectors contain the raw temperature values shown in Figure 2 of the main text; the other vectors contain temperature values that are shifted to have mean zero and scaled to have variance one.

\item {\tt logit\char`_p\char`_LBs.txt}

This file contains the three variables {\tt hot\char`_limits},  {\tt cold\char`_limits}, and {\tt wet\char`_limits}, as well as a {\tt region} variable. Each of the {\tt limits} variables contain the lower bound for the logit probabilities used in the model fitting for the main text; these represent the $-L$ value introduced in Supplementary Section \ref{priorSpec}.

\end{enumerate}

\subsection{Primary functions}

The replication code file contains sample implementations of each of the following functions. Default values of the inputs are indicated.

First, a wrapper function to run the componentwise interweaving MCMC sampler is:

{\singlespacing
\begin{verbatim}
timeRR_month( zA_mat, zN_mat, n_sims, gmt_A, gmt_N, 
  ITER = 10000, start_keep = 1, n_tune = 6, 
  n_iter_tune = c(rep(400,3),rep(800,3)),
  thin = 1, event_type, alpha_delta_corr_prop = -0.98,
  betaA_corr_prop = 0, betaN_corr_prop = 0, 
  logit_p_UB = 15, logit_p_LB = -15 )
\end{verbatim}
}

\noindent The variables {\tt zA\char`_mat} and {\tt zN\char`_mat} must be $M\times12$ matrices of monthly count variables (where $M$ is the total number of years) for the ALL and NAT scenarios, respectively. The variables {\tt n\char`_sims}, {\tt gmt\char`_A}, and {\tt gmt\char`_N} require vectors of length $M$, which contain the ensemble size, ALL scenario global temperature value, and NAT scenario global temperature value, respectively, for each year. The next five variables specify the run length of the MCMC: {\tt ITER} is the number of post-tuning MCMC iterations; {\tt start\char`_keep} (between 1 and {\tt ITER}) is the number of burn-in iterations that are discarded; {\tt thin} specifies if the saved MCMC output is thinned ({\tt thin} $> 1$) or not ({\tt thin} $= 1$); {\tt n\char`_tune} is the number of MCMC cycles used to tune the RWMH standard deviations, for {\tt n\char`_iter\char`_tune} iterations per tune cycle. The variable {\tt event\char`_type} must be one of {\tt "hot"}, {\tt "cold"}, or {\tt "wet"}; the remaining variables specify the proposal correlations and limits on the logit probabilities.

Next, two functions are provided to summarize the MCMC output. These are

{\singlespacing
\begin{verbatim}
calculate_probs_RR( fit_obj, years = NULL, lp = 0.025, 
  up = 0.975 )
\end{verbatim}
}

\noindent and

{\singlespacing
\begin{verbatim}
exceed_cutoff_RR( fit_obj, gmt_A_ref, gmt_N_ref, cutoff, 
  direction = ">", lp = 0.025, up = 0.975, years = NULL )
\end{verbatim}
}

\noindent The function {\tt calculate\char`_probs\char`_RR} takes the output of {\tt timeRR\char`_month} (using the {\tt fit\char`_obj} variable) and calculates the yearly probabilities ($p_{kt}$) and risk ratio ($RR_t$), as well as provides a point estimate (the posterior median) and a credible interval estimate for each probability and risk ratio. The {\tt lp} and {\tt up} variables allow for the lower and upper quantiles for the credible interval, respectively. The {\tt years} variable allows for an optional specification of what years are represented by the data.

The {\tt exceed\char`_cutoff\char`_RR} function calculates the adjusted risk ratio ($\widetilde{RR}_t$) as well as the exceedance probabilities $\pi$. Again, the function takes the output of {\tt timeRR\char`_month} (using the {\tt fit\char`_obj} variable) and adjusts the corresponding risk ratio to behave as if the arose from a stationary climate where the global temperature for ALL and NAT were {\tt gmt\char`_A\char`_ref} and {\tt gmt\char`_N\char`_ref}, respectively. The {\tt cutoff} variable is either a scalar or vector of length two, indicating a single cutoff for which to compare the adjusted risk ratio against or two values for which to determine the probability of the adjusted risk ratio lying between the two values. The {\tt direction} variable must be one of {\tt ">"}, {\tt "<"}, or {\tt "<>"}, to indicate whether the exceedance probability should be greater than, less than, or between the cutoff value(s) given. The {\tt lp}, {\tt up}, and {\tt years} variables are as in {\tt calculate\char`_probs\char`_RR}.

Finally, a function is provided to plot either the $p_{kt}$ or the $RR_t$:

{\singlespacing
\begin{verbatim}
plot_over_time( m, ub, lb, year, n, x_tic = 3, 
  grp_name = NULL, grp_pal = brewer.pal(3, "Dark2"), 
  title_txt = NULL, ylab = NULL, xlab = NULL,  
  y_lb = min(lb), y_ub = max(ub), 
  plot_p = TRUE, rr_bounds = c(-3,5,0.5,3) )
\end{verbatim}
}

\noindent The {\tt m}, {\tt ub}, and {\tt lb} variables take vectors of length $M$ of, respectively, the point estimate, credible interval upper bound, and credible interval lower bound for each year of the variable to be plotted. The variable {\tt year} contains a vector of which years are included in the analysis; {\tt n} is a vector of the ensemble sizes for each year (for the legend); and {\tt x\char`_tic} indicates the spacing between the years plotted on the $x$-axis. Next, {\tt grp\char`_name} is the name of the legend, and {\tt grp\char`_pal} indicates the color palette to use ({\tt brewer.pal} is a function from the {\tt RColorBrewer} package). The options {\tt title\char`_text}, {\tt ylab}, {\tt xlab}, {\tt y\char`_lb}, and {\tt y\char`_ub} define standard plotting parameters. Finally, {\tt plot\char`_p} indicates whether the probabilities $p_{kt}$ are to be plotted ({\tt TRUE}), or if instead the risk ratio $RR_t$ are to be plotted ({\tt FALSE}), and {\tt rr\char`_bounds} is a vector of length four which contains, respectively, the exponent of 10 for the lower and upper bounds of the $y$-axis, the spacing between labels, and how many decimal places to round the labels.

\end{appendix}

\end{document}